\documentclass[twocolumn]{aastex63}

\usepackage{amsmath}
\usepackage{xspace}
\usepackage{enumitem}
\usepackage[frozencache,cachedir=.]{minted}
\usepackage{lineno}
\usepackage{xcolor}

\newcommand{\ttt}[1]{\texttt{#1}}

\newcommand\Abacus{\textsc{Abacus}\xspace}
\newcommand\AbacusSummit{\textsc{AbacusSummit}\xspace}

\newcommand{\hMpc}{\ensuremath{\mathit{h}^{-1}\ \mathrm{Mpc}}}
\newcommand{\hGpc}{\ensuremath{\mathit{h}^{-1}\ \mathrm{Gpc}}}
\newcommand{\hkpc}{\ensuremath{\mathit{h}^{-1}\ \mathrm{kpc}}}

\newcommand{\hMsun}{\ensuremath{\mathit{h}^{-1}\ \mathrm{M_\odot}}}
\newcommand{\Msun}{\rm\,M_\odot}
\newcommand{\LCDM}{$\Lambda$CDM\xspace}

\newcommand{\bfr}{{\bf r}}

\newcommand{\bfF}{{\bf F}}

\newcommand\code[1]{\texttt{#1}}

\shorttitle{The \textsc{Abacus} $N$-body Code}
\shortauthors{Garrison et al.}
\graphicspath{{./}{figures/}}

\begin{document}

\title{The \textsc{Abacus} Cosmological $N$-body Code}

\correspondingauthor{Lehman Garrison}
\email{lgarrison@flatironinstitute.org}

\author[0000-0002-9853-5673]{Lehman H. Garrison}
\affiliation{Center for Computational Astrophysics, Flatiron Institute \\
Simons Foundation, 162 Fifth Ave.\\
New York, NY 10010, USA}

\author{Daniel J. Eisenstein}
\author{Douglas Ferrer}
\author{Nina A. Maksimova}
\affiliation{Center for Astrophysics $|$ Harvard \& Smithsonian \\
60 Garden St, Cambridge, MA 02138, USA}

\author{Philip A. Pinto}
\affiliation{Steward Observatory, University of Arizona, 933 N. Cherry Ave., Tucson, AZ 85121}



\begin{abstract}
We present \textsc{Abacus}, a fast and accurate cosmological $N$-body code based on a new method for calculating the gravitational potential from a static multipole mesh.  The method analytically separates the near- and far-field forces, reducing the former to direct $1/r^2$ summation and the latter to a discrete convolution over multipoles.  The method achieves 70 million particle updates per second per node of the Summit supercomputer, while maintaining a median fractional force error of $10^{-5}$.  We express the simulation time step as an event-driven ``pipeline'', incorporating asynchronous events such as completion of co-processor work, Input/Output, and network communication.  \textsc{Abacus} has been used to produce the largest suite of $N$-body simulations to date, the \textsc{AbacusSummit} suite of 60 trillion particles (Maksimova et al., 2021), incorporating on-the-fly halo finding.  \textsc{Abacus} enables the production of mock catalogs of the volume and resolution required by the coming generation of cosmological surveys.
\end{abstract}



\section{Introduction} \label{sec:intro}
Cosmological $N$-body simulations trace the gravitational dynamics of matter on large scales.  By sampling the phase-space distribution of matter with $N$ discrete particles, $N$-body simulations evolve the particles under mutual self-gravity to trace the emergence of structure.  The densest regions are gravitationally bound structures called ``halos'', strung together by filaments and sheets, forming a scaffolding upon which statistical realizations of the luminous galaxy field can be painted.  Thus, by simulating an $N$-body system for a given cosmology, one can constrain cosmological models by comparing the clustering properties of the simulation to that of observed galaxies.

The increased statistical power of modern galaxy surveys (e.g.~the Dark Energy Spectroscopic Instrument, DESI, \citealt{Levi+2013}; \textit{Euclid}, \citealt{Laureijs+2011}; \textit{Roman Space Telescope}, \citealt{Spergel+2015}; the Vera C.\ Rubin Observatory, \citealt{Ivezic+2019}) has demanded similarly increased levels of precision in the theoretical predictions of galaxy clustering.  Indeed, one wants to show that the bias of a given analysis method is many times smaller than the statistical power of the survey, not just equal to it.  Therefore, simulation datasets with many times the statistical power of the survey itself must be constructed, lest the analysis be declared systematics-limited when the truth might be much better.

This demand for large volumes is compounded by the need to resolve the low-mass halos that hold faint, low-mass galaxies.  Surveys like DESI will target emission-line galaxies (ELGs) between redshifts 0.6 and 1.6 over 10s of Gpc$^3$; ELGs inhabit halos down to masses of $1\times 10^{11} M_\sun$ \citep{Gonzalez-Perez+2018,Avila+2020}, requiring a particle mass of $2\times 10^9 M_\sun$ for a minimal sampling of 50 particles per halo.  Such halos are also smaller in physical extent and therefore demand higher force resolution (i.e.~smaller softening length), which in turn demands smaller time steps.  Therefore, the current ``flagship'' $N$-body simulations all contain hundreds of billions to trillions of particles and are run at major supercomputer facilities.  Such simulations include the \textit{Euclid} Flagship simulations of the \textsc{PKDGrav3} code \citep{Potter+2017}, the Outer Rim simulation of the \textsc{HACC} code \citep{Heitmann+2019}, the Uchuu simulations of the \textsc{GreeM} code \citep{Ishiyama+2020}, and the \AbacusSummit simulations of the \Abacus code \citep{Maksimova+2021}.

$N$-body simulations are computationally challenging.  The force evaluation is expensive due to the large number of particles and the ``all to all'' nature of the long-range gravitational force.  Therefore, nearly all codes split the gravitational force kernel into near-field and far-field components \citep[e.g.][]{Efstathiou+1985,Springel_2005}.  The far-field can employ distant-source approximations like multipole expansion or particle mesh and is commonly handled in Fourier space.  This also presents the opportunity to apply periodic boundary conditions, thereby rendering the simulation homogeneous.  The near field may similarly be accelerated with approximations like the Barnes-Hut tree \citep{Barnes_Hut_1986}.

Even with fast force evaluation, a large amount of memory is required to hold the state.  Large simulations therefore require distributed memory systems, complicating the implementation of simulation codes (but see \Abacus's disk buffering mode below, or the compression techniques of \textsc{CUBE}, \citealt{Yu+2018}).  This is one motivation for the use of second-order leapfrog integration \citep{Quinn+1997}, rather than higher-order methods that require additional force evaluations and copies of the state (and are generally not symplectic as well).

``Approximate'' $N$-body methods have made great progress in recent years, achieving good accuracy on intermediate to large scales.  Simulation techniques now exist on a spectrum of performance and accuracy, including methods like COLA \citep{Tassev+2013} and its variants; \textsc{FastPM} \citep{Feng+2016}; and \textsc{GLAM} \citep{Klypin_Prada_2018}.  In complicated analyses involving high-order correlation functions and coupling of small to large scales, however, as with redshift-space distortions, it remains useful to have full $N$-body result.

In short, $N$-body simulations are useful but expensive, so developing codes that can run them quickly but accurately is an important program for large-scale structure cosmology.  The \Abacus code was conceived in \cite{Metchnik_2009} in this context.  The central pillar of \Abacus is the mathematical advance in the computation of the far-field force from a convolution of a mesh of multipoles, with an analytic separation of the near- and far-field forces.  But the structure of the mathematical computation suggests opportunities to structure the code for high performance, which we will discuss throughout this work.

Writing high-performance code also means adapting to modern computer hardware (unless one is to procure custom hardware, as with the GRAPE machines, \citealt{Makino_Daisaka_2012}).  We will discuss how the \Abacus code is adapted to run on modern systems with technologies like superscalar cores and co-processing with graphic processing units (GPUs).

This paper is one in a series about \Abacus being published alongside the release of the \AbacusSummit suite of simulations.  The present work describes the code and basic validation: data structures, algorithms, interfaces, optimizations, and force and time step accuracy.  \cite{Maksimova+2021} describes the \AbacusSummit simulations and performance on the Summit supercomputer.  \cite{Hadzhiyska+2021} describes the on-the-fly CompaSO halo finder.  \cite{Bose+2021} describes the construction of merger trees.  Pinto et al.~(in prep.) describes the mathematical methods.

Past published work on \Abacus includes initial conditions \citep{Garrison+2016}, the Abacus Cosmos suite \citep{Garrison+2018}, and a realization of the \textit{Euclid} code comparison simulation \citep{Garrison+2019}.

The rest of this paper is laid out as follows.  In Section \ref{sec:overview}, we introduce \Abacus, the force solver, and the organization of the computation.  In Section \ref{sec:dynamical_evolution}, we discuss the time stepping and detail the force computation, followed by accuracy tests.  In Section \ref{sec:pipeline}, we demonstrate the organization of the computation into a slab pipeline, and detail some low-level aspects of the memory, data, and thread management in Section \ref{sec:memory_and_threads}.  In Section \ref{sec:group}, we discuss our on-the-fly group finder, and in Section \ref{sec:toplevel}, we present some of the software surrounding \Abacus proper.  We discuss the use of commodity hardware to run \Abacus simulations in Section \ref{sec:hardware}, and catalog some notable simulations executed with \Abacus in Section \ref{sec:simulations}.  We summarize in Section \ref{sec:summary}.

\section{Overview}\label{sec:overview}
\subsection{Force Solver}\label{sec:force_solver}

\begin{figure}[ht]
\begin{center}
\includegraphics[width=1.0\columnwidth]{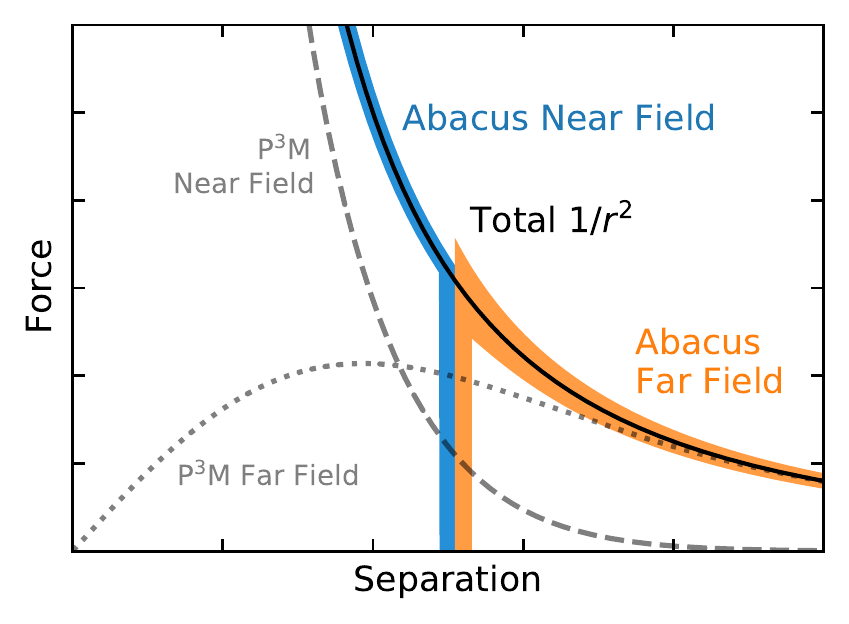}
\caption{A schematic illustration of the exact near-field/far-field separation of the \Abacus force computation.  The grey lines represent other schemes, like particle mesh, in which the far field (dotted line) overlaps the near field, which requires compensation by the near field (dashed line).  In the \Abacus force computation (shaded lines), the entire $1/r^2$ force is given by exactly one of the near field and far field.  The increasing width of the \Abacus far-field near the transition shows schematically that the far-field is less accurate at smaller separations.  The actual transition error is small.
\label{fig:force_split}}
\end{center}
\end{figure}

\begin{figure}[ht]
\begin{center}
\includegraphics[width=1.0\columnwidth]{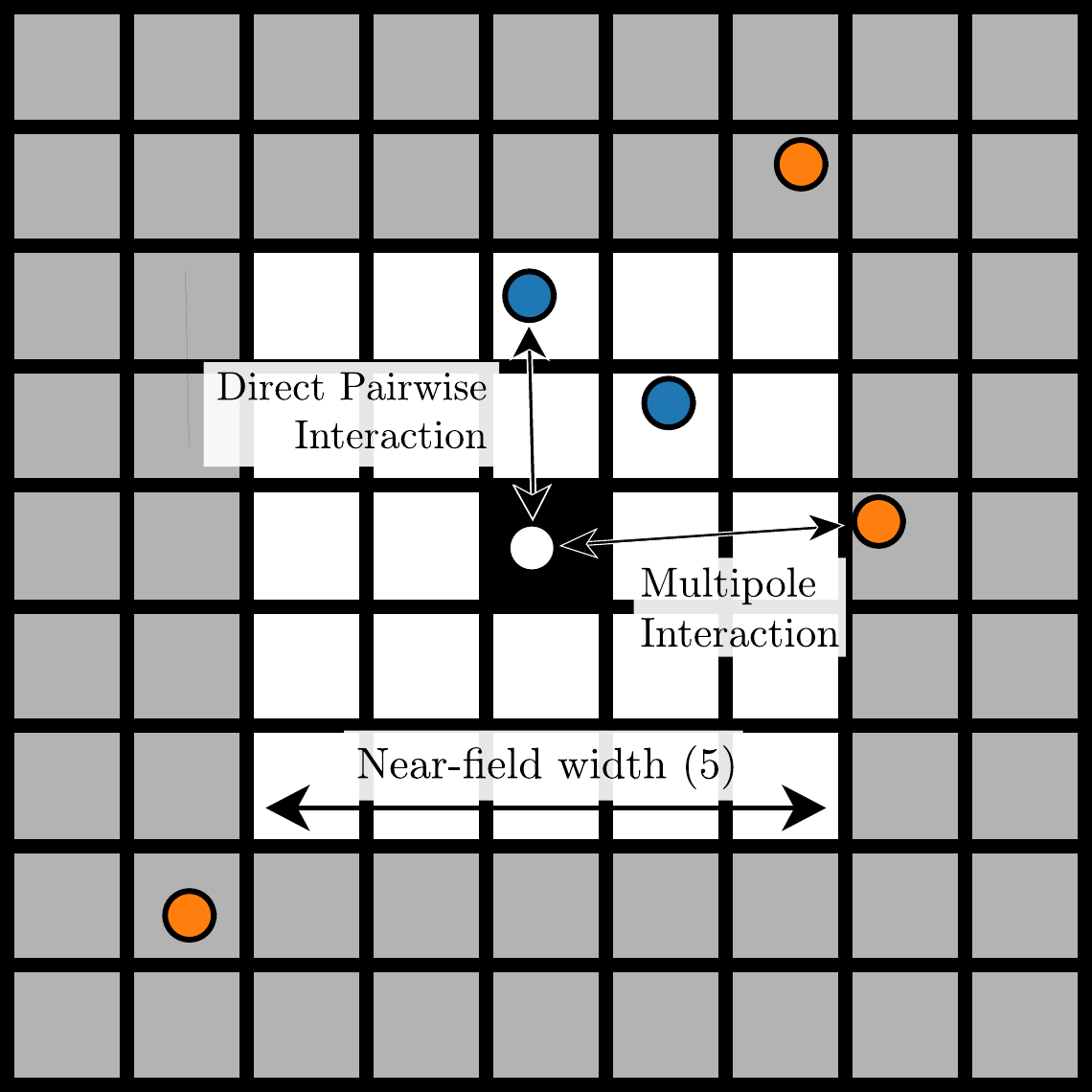}
\caption{The \Abacus domain decomposition.  Forces on the particle in the black cell may come from the near field (white cells and black cell) or far field (shaded cells).  Particles in the near field interact via direct $1/r^2$ Newtonian gravity; particles in the far field use a high-order multipole approximation to same.  The far-field acceleration in the black cell derives from a Taylor expansion of the potential about the cell center and is specific to that cell.
\label{fig:domain_decomp}}
\end{center}
\end{figure}

\Abacus is a high-performance code for massive cosmological $N$-body simulations with high force accuracy.
It is based on a novel method for solving the Poisson equation, developed in \citet{Metchnik_2009}, that features an exact decomposition of the near-field and far-field force.  In other words, each pairwise interaction is given by either the near field or the far field, not both, as illustrated in Figure \ref{fig:force_split}.  The far-field computation is based on the multipole expansion, in which the Taylor series expansion up to polynomial order $p$ of the gravitational potential around one location is related via linear coefficients to the multipole moments to order $p$ of the density field in a distant region centered on a second point.  The linear coefficients depend on the separation of the two expansion points, but can be precomputed.  The key advance in the method of \citet{Metchnik_2009} is to decompose the simulation in a 3-d Cartesian grid of cells, computing the multipole moments in each cell.  To then compute the Taylor series of the potential in a particular target cell involves summing over the contributions from the multipoles from all sufficiently distant cells (including all periodic copies).  But because of the discrete translation symmetry of the Cartesian grid, one can perform this summation for all target cells at once as a convolution using discrete Fourier transforms.  In other words, we compute the Fourier transforms of the grid of the multipoles, compute the linear combinations with the Fourier transforms of the grid of all of the weights, and then inverse Fourier transform to yield the grid of all of the Taylor series coefficients.

With the far-field computation handling all of the sufficiently distanced cells, as illustrated in Figure \ref{fig:domain_decomp}, the near-field force is computed as the Newtonian gravity from all of the close cells, using open boundary conditions and the very simple softened $1/r^2$ force.  This is not an approximation, since the far field includes all periodic images of the close cells.
The simplicity of the near-field force kernel and the repeated geometric structure of the near-field domain lends itself very well to co-processor acceleration with GPUs.  The resulting total forces are highly accurate, with only one parameter, the order of the multipole expansion, controlling the accuracy of the entire method. We will show that a modest order, $p=8$, and keeping only $5^3$ neighboring cells in the near field, is sufficient to give excellent force accuracy for cosmological simulations.  Remarkably, the Fourier-space convolution of the far field, the computation of the multipole moments, and the evaluations of the Taylor series of the potential can be computed quickly enough to keep up with the tremendous speed of modern GPU computing.

\subsection{Two-Stage Execution}\label{sec:twostage}

Because the grid is fundamental to the force computation, the \Abacus data model and work flow are heavily organized around its grid.  The grid is cubic of size $K^3$ (with $K$ sometimes labeled as Cells Per Dimension or CPD).  We typically choose $K$ so that the average cell contains 30--100 particles.  Cells are organized into planar \textit{slabs} that are one cell wide in a chosen direction, $x$.  Particles and other properties are stored in grid order, indexed by slab and by cell.  It is not necessary that the grid be commensurate with the size of the particle lattice (in the case that the initial conditions are generated by perturbing such a lattice).

The next major concept in \Abacus is to note that the near-field/far-field split ensures that the domain of the near-field computation is strictly bounded.  Indeed, all aspects of the processing save for the Fourier transforms in the far field computation require only a localized view of the data, spanning a few cells in each direction.  We use this property to minimize memory consumption by organizing the calculations into a thin 1-dimensional sweep of the computational domain, which we call the {\it slab pipeline}.  Slabs of particle data (positions \& velocities) and cell-based Taylor series coefficients enter the pipeline and go through multiple computational steps, resulting in updated particle data and cell-based multipole moments when they exit the pipeline.  Any extra memory used in these computations need only be held for the slabs that are in the pipeline.  After sweeping the entire volume to update the particles, we then do the far-field operation to convert the multipole moments into Taylor series coefficients, readying the system for the next sweep.  In this manner, 
the \Abacus time step operates in a ``tick-tock'' fashion: a local, planar sweep through the volume that loads each particle once, and a global step that need not touch any particles, merely execute a Fourier-space convolution on the cell grid of the multipoles---a far smaller amount of data.

The well-ordered sweep of the particle data offers a further opportunity to stream this data from disk, as this contiguous I/O on large files is amenable to inexpensive arrays of spinning hard drives (Sec.~\ref{sec:hardware}).  A single node with such disk can therefore execute a simulation larger than fits in memory---a so-called \textit{out-of-core algorithm}.  However, when enough memory is available, such as when employing a cluster of nodes, it can be advantageous to hold the entire particle set in memory.  In this case, we employ Linux ramdisk (Sec.~\ref{sec:ramdisk}).  \Abacus therefore has no special hardware requirements, but can instead operate in one of two modes---disk I/O or ramdisk, depending on available hardware.

The locality of the slab pipeline benefits the parallel implementation of \Abacus, too.  To a given node, it does not matter if the not-in-memory data is on disk or on another node.  Each node can hold a range of slabs, and slabs can be passed once per time step in a toroidal transfer---no tightly-coupled inter-node communication or ghost (padding) zones are required.  The far-field requires one distributed Fourier transform---an all-to-all operation---per time step.  The parallel implementation is discussed in Section~\ref{sec:parallel}.

\subsection{Intuition: Abacus Compared to PM and Tree Methods}
The central advance of \Abacus is the exact near-field/far-field split.  To understand how \Abacus excludes the near-field region from the far-field force kernel, a tempting comparison is particle-mesh based methods that also include a short-range force component, like Particle-Particle Particle-Mesh (P$^3$M).  However, the PM part of these calculations deposits the particle mass onto a monolithic mesh on which the long-range force is calculated, offering neighboring particles no opportunity to ``exclude'' each other, aside from a post-facto cut in Fourier space.  \Abacus never bins particles on a mesh, instead computing multipole moments in cells.  To see how this helps, though, a more instructive comparison may be a tree method.

In a tree, or hierarchical, method such as Barnes-Hut \citep{Barnes_Hut_1986} or the Fast Multipole Method \citep[FMM,][for a recent $N$-body implementation]{Greengard_Rokhlin_1987,Potter+2017}, the domain is decomposed into hierarchical cells in which multipole moments are computed.  A given cell is free to interact with another cell directly, computing all pairwise interactions, or with its multipoles.  If we consider direct interaction ``near field'' and multipole interaction ``far field'', then we can say that a tree method has an exact near-field/far-field split.  \Abacus works the same way, with the near field given by direct interaction and the far field by multipoles, except that multipoles are computed on a Cartesian grid instead of a tree.  This rigid structuring of the cells allows us to re-phrase the problem as a convolution over cells instead of many separate interactions of pairs of cells.  The convolution is amenable to acceleration with FFT methods which also offers a chance to include periodic boundary conditions for ``free''.

PM methods additionally benefit from smaller cell size, while the force accuracy of \Abacus is largely independent of the number of cells, because the accuracy of the multipole forces is set by the opening angle to the far-field, which is independent of grid size.

\section{Dynamical Evolution}\label{sec:dynamical_evolution}
\subsection{Time Stepping}\label{sec:timestep}
The \Abacus particle update cycle is the standard second-order leapfrog Kick-Drift-Kick \citep[KDK;][]{Quinn+1997} using a global time step.  The time step size $\Delta a$ is chosen at the beginning of each time step using the ratio of the root-mean-square velocity in the cell to the maximum acceleration in that cell.  The quantities are computed in each cell, and the most conservative cell sets the global time step, scaled by time step parameter $\eta_\mathrm{acc}$ and a factor of $aH$ to yield $\Delta a$ from $\Delta t$.  Additionally, a global rms velocity to maximum acceleration ratio is computed and is used if it is larger than the cell-wise criterion to guard against abnormally cold cells causing catastrophically small time steps.  Finally, the time step is limited to a fraction $\eta_\mathrm{H}$ of the Hubble time.  This only takes effect at early redshift, before the first structures form.  We typically choose $\eta_\mathrm{H}$ to require about 33 steps per $e$-fold of the scale factor.  All together, the time step criterion is
\begin{equation}\label{eqn:timestep}
\begin{split}
    \Delta a = &\min\Bigg( \eta_\mathrm{H} a, \\
    & \eta_\mathrm{acc}aH\max\left(\min_\mathrm{c \in cells}\left[\frac{v_\mathrm{rms}}{a_\mathrm{max}}\right]_c, \frac{v_\mathrm{rms,global}}{a_\mathrm{max,global}}\right) \Bigg).
\end{split}
\end{equation}

The $v_\mathrm{rms}$ is computed without subtracting off the mean velocity in cells, and therefore it is not formally Galilean invariant.  This implies that the time step determination, which seeks low $v_\mathrm{rms}$ and high $a_\mathrm{max}$, will primarily arise from large halos in low-velocity flows.  This is desirable, as it ensures that the time step requirements of similar halos in high-velocity regions are not underestimated. Regardless, this modulation is small, as RMS velocities of large halos are much larger than typical bulk flows.  Furthermore, as the time step is applied globally, the mean effect of the Galilean invariance will be absorbed by tuning $\eta_\mathrm{acc}$.

The time step is also foreshortened to land on \textit{full time slice} output redshifts (further described in Sec.~\ref{sec:outputs}).  Other output types, like \textit{subsample time slices}, are ``best effort'', meaning the time step is not shortened to land on an exact redshift.  This can save dozens of time steps when there are dozens of subsample time slices.

Since the global time step is tied to the most stringent time step criterion in the whole simulation volume, the inaccuracy from the finite time step will be felt primarily in the densest regions, typically the cores of the largest halos.  
In the \AbacusSummit simulations, for example, we typically take $1100$ time steps with particle mass $2\times 10^9\ \hMsun$ and $7\hkpc$ softening, while similar simulations with sub-cycling take up to a few times more steps at the finest level of cycling \citep{Potter+2017}.  However, we stress that most regions of the simulation have a more favorable ratio of the time step to the local dynamical time and hence more accurate integrations.

\begin{figure}
    \centering
    \includegraphics[width=\columnwidth]{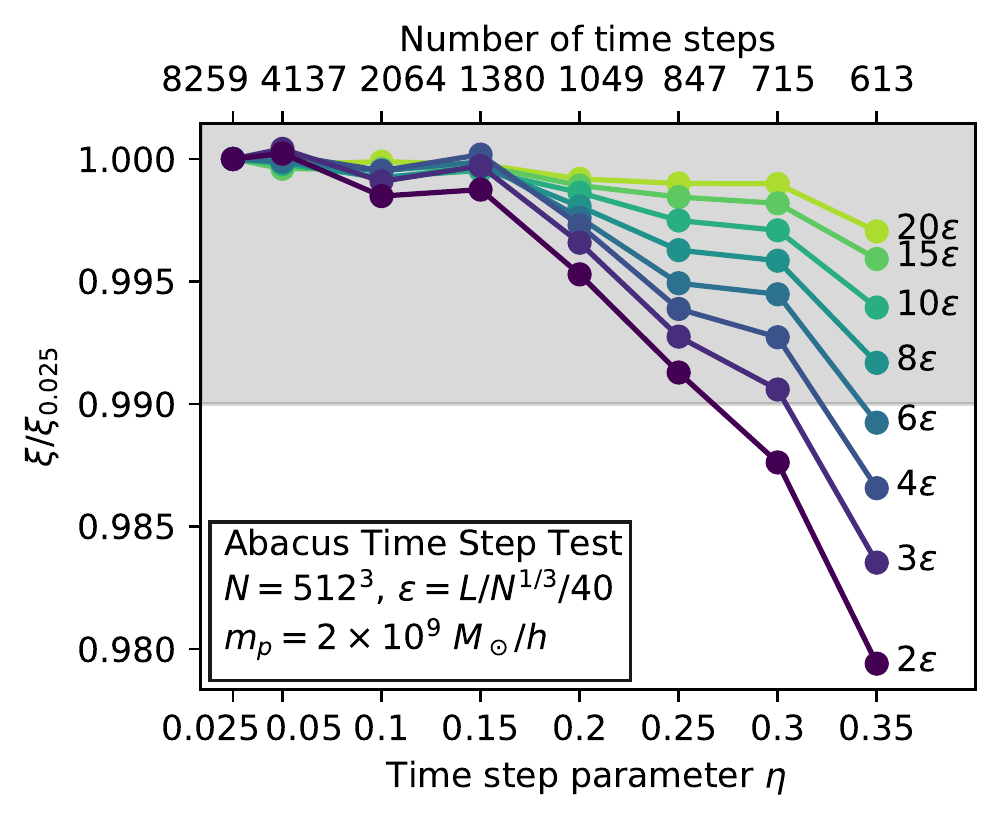}
    \caption{Convergence of the small-scale correlation function with respect to time step parameter $\eta$, or \ttt{TimeStepAccel}, for different spatial scales. Spatial scales ($2\epsilon$, $3\epsilon$, etc) are labeled as multiples of the Plummer-equivalent softening length $\epsilon$.  The mapping of $\eta$ (bottom axis) to number of time steps (top axis) depends on the simulation; variations in volume or mass resolution will affect the halo dynamical times. \AbacusSummit used $\eta=0.25$.}
    \label{fig:timestep}
\end{figure}

In Figure~\ref{fig:timestep}, we measure the small-scale 2PCF as a function of time step, and see that it is missing about 1\% of power on scales twice the softening length for $\eta = 0.25$ (the value used in \AbacusSummit), using an 8000 step simulation as a reference.  While not a direct measure of halo structure, the small-scale 2PCF is dominated by the densest regions of the simulation; namely, halo cores.  Applications that require accurate halo profiles should evaluate their convergence with respect to time step, but for applications that marginalize over the details of innermost halo structure, we expect $\mathcal{O}(1000)$ time steps to be sufficient.  We note that this number of time steps would likely not be sufficient if the time step size were fixed in $a$ or $\log a$, as too many time steps would be spent in early epochs where dynamical times are long.

Additionally, the time step is not necessarily the dominant systematic error on small scales.  Softening and finite mass resolution necessarily impose a UV cutoff that prevents the simulation from reaching the continuum limit.  \cite{Joyce+2020} demonstrates this in the context of an $n=-2$ scale-free simulation, where the finite mass resolution is shown to place a lower bound on the error of about 5\% at $3\epsilon$ to $6\epsilon$ for epochs typical of \LCDM simulations---an effect an order of magnitude larger than the $<0.5\%$ time step error for $\eta \le 0.2$ in Fig.~\ref{fig:timestep}.  \cite{Garrison+2021} use additional scale-free simulations to demonstrate how to tune the softening length to yield a similar error budget to the mass resolution (and show that such fixing the softening in proper coordinates uses fewer time steps at fixed error budget).  Both of these effects---softening and mass resolution---suggest that \Abacus simulations are not limited in their accuracy primarily by time step.

Aside from the softening length, the functional form of the force softening law impacts the accuracy of the small-scale clustering as well.  We turn to this next.

\subsection{Force Softening}\label{sec:softening}
As is well-known, it is desirable in the $N$-body problem to use a softened force law to avoid the very large accelerations that result when two particles are unusually close together.
A common version is Plummer softening \citep{Plummer_1911}, where the $\bfF(\bfr) = \bfr/r^3$ force law is modified as
\begin{equation}\label{eqn:plummer}
\bfF(\bfr) = \frac{\bfr}{(r^2 + \epsilon_p^2)^{3/2}},
\end{equation}
where $\epsilon_p$ is the softening length.  This softening is fast to compute but is not compact, meaning it never explicitly switches to the exact form at any radius, only converging quadratically to $r^{-2}$. This affects the growth of structure on scales many times larger than $\epsilon_p$, as demonstrated in \cite{Garrison+2019}.  Formally, this is inconsistent with the Abacus far-field model, which uses the $1/r^2$ force law.

\begin{figure}
    \centering
    \includegraphics[width=\columnwidth]{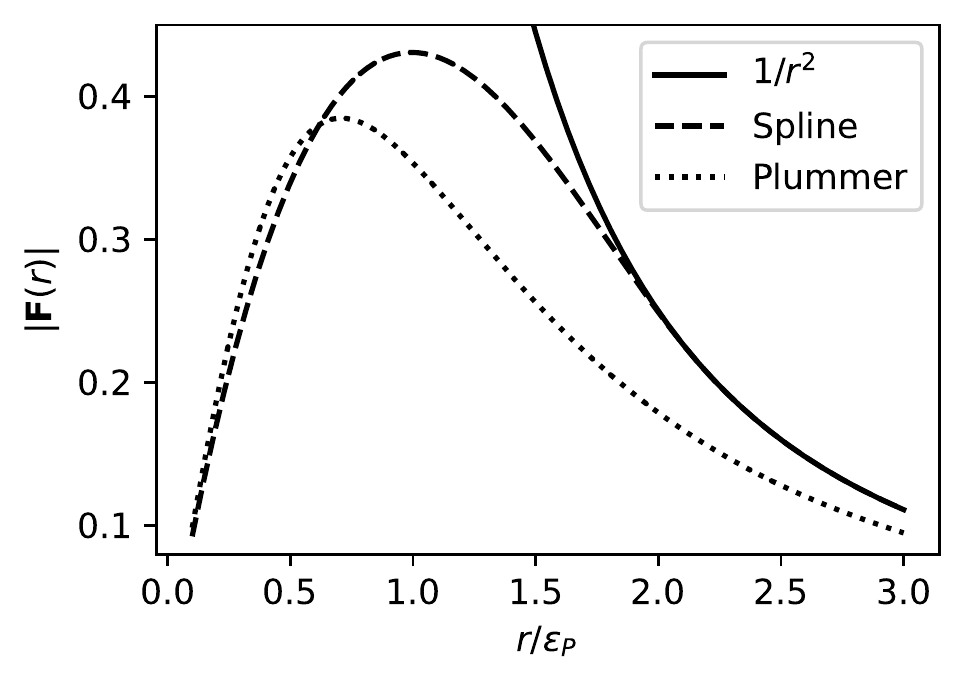}
    \caption{The \Abacus spline force softening (Eq.~\ref{eqn:spline}), compared with Plummer softening (Eq.~\ref{eqn:plummer}) and the $1/r^2$ force. The \Abacus spline softening length is $2.16\epsilon_p$ to match the small-scale Plummer force and hence the small-scale orbital time.  The spline form matches the exact form at smaller radii than Plummer softening.}
    \label{fig:softening}
\end{figure}

Spline softening is an alternative in which the force law is softened for small radii but explicitly switches to the unsoftened form at large radii.  Traditional spline implementations split the force law into three or more piecewise segments \citep[e.g.~the cubic spline of][]{Hernquist_Katz_1989}; we split only once for computational efficiency and to avoid code path branching.
We derive our spline implementation by considering a Taylor expansion in $r$ of Plummer softening (Eq.~\ref{eqn:plummer}) and requiring a smooth transition at the softening scale up to the second derivative\footnote{A Taylor expansion in $r^2$ is also possible, but we discard that solution due to a large plateau of constant angular frequency near $r\sim 0$ that could potentially excite dynamical instabilities.}.  This yields
\begin{equation}\label{eqn:spline}
\bfF(\bfr) =
\begin{cases}
\left[10 - 15(r/\epsilon_s) + 6(r/\epsilon_s)^2\right]\bfr/\epsilon_s^3, & r < \epsilon_s; \\
\bfr/r^3, & r \ge \epsilon_s.
\end{cases}
\end{equation}
This was first presented in \cite{Garrison+2016} and is plotted alongside Plummer softening and the exact $1/r^2$ force law in Figure~\ref{fig:softening}.

The softening scales $\epsilon_s$ and $\epsilon_p$ imply different minimum dynamical times---an important property, as this sets the step size necessary to resolve orbits.  We always choose the softening length as if it were a Plummer softening and then internally convert to a softening length that gives the same minimum pairwise orbital time for the chosen softening method.  For our spline, the conversion is $\epsilon_s = 2.16\epsilon_p$.

For all reasonable cases in \Abacus, we choose values of the softening length that imply that the spline force returns to the exact form well before the near-field/far-field transition, so as to avoid any discontinuity in the pairwise force at that boundary. 

\subsection{Far-field Computation}
\subsubsection{Overview}
We now turn to the details of the computation of the force, starting with the far field.  As introduced in Sec.~\ref{sec:force_solver}, the far field employs a multipole method that relates the order $p$ multipole expansion in one cell to the Taylor series expansion of the gravitational potential in all other cells via a set of linear coefficients.  Because of the discrete translation symmetry of the $K^3$ cell grid, the total Taylor series coefficients in all cells can be computed simultaneously from the multipole coefficients of all cells via a discrete convolution on the grid.  This convolution can be implemented efficiently in Fourier space.

The set of linear coefficients is called the \textit{derivatives tensor}, or simply the \textit{derivatives}, as it arises from the derivative of the gravitational potential.  Convolving the derivatives with the multipoles yields the Taylor series coefficients, shortened to the \textit{Taylors}.  Since this convolution is executed in Fourier space, it requires a global view of the box: at least one full dimension must be in memory at a time so that the discrete Fourier transform can be executed along that dimension.  During the slab pipeline, the $y$ and $z$ dimensions are in memory, but not the $x$ (the slab dimension).  Therefore, the computation proceeds as follows.  As the slab pipeline executes its planar sweep through the volume, it produces $K$ slabs of cell multipoles, computed on the updated (drifted) particle positions.  Each slab is Fourier transformed in two dimensions and, when using the out-of-core algorithm (introduced in Sec.~\ref{sec:twostage}), can be saved to persistent storage so that it can be discarded from memory.  Then, when the slab pipeline has completed and all $K$ multipole slabs have been stored, a small range is read from each of the $K$ slabs, small enough that we can fit all $K$ in memory, but large enough to overcome the latency of the read.  From these data, we can form a series of individual pencils spanning $x$ at constant $y$ and $z$.  Each pencil is Fourier transformed and the derivatives tensor is applied, followed by the inverse FFT in $x$.  The result is written back to persistent storage and repeated, building up slabs of Taylors ready for reading by the pipeline in the next time step.  The inverse $yz$-FFT is applied by the pipeline after reading the Taylors slab.  

Every time step thus consists of two sub-steps: the slab pipeline and the cross-slab \textit{convolution}---the ``tick-tock'' execution introduced in Section \ref{sec:twostage}.



The derivatives are fixed for a given grid size $K$, near-field radius $R$, and multipole order $p$, and are precomputed in a small amount of time.  They include the contribution from all periodic images of the box, including the images of the central and near-field cells. Therefore, while each pairwise interaction between two particles is given by either the near-field or far-field, any two particles will have multiple pairwise interactions due to the multiple periodic images.  The total force is equivalent to Ewald summation \citep{Ewald_1921}, up to finite multipole order.

We note that while the transformation from multipoles to Taylors is linear, it does involve many terms, scaling as $O(p^5)$ \citep{Metchnik_2009}.  We use vector operations to speed this work (Section \ref{sec:multipoles_simd}), as well as numerous symmetries.  For example, the derivative terms obey predictable even or odd parity, so that their Fourier transforms are either purely real or purely imaginary.  The product with the generally complex multipole coefficients can therefore be coded as a series of real times complex products, halving the work compared to the naive view.

The multipole order $p$ controls the accuracy of the force evaluation.  Order $p=8$ is our usual choice, which skews towards accuracy at the cost of performance.  We present several accuracy tests in Section \ref{sec:force_accuracy} on random particle distributions, lattices, and \LCDM simulations, the summary of which is that $p=8$ is exceptionally accurate, with typical fractional errors of order $10^{-5}$.

One reason we choose the multipole order to give such high force accuracy is that our domain decomposition is a structured mesh.  When computing forces on such a repeating structure, the force error patterns are likely to not be homogeneous and random: they will vary based on position in the cell and approximately repeat in every cell.  Such a spatially repeating error could readily appear clustering statistics, which are some of the primary quantities we wish to measure from these simulations.  These errors are not inherently larger than the errors from a dynamic or unstructured decomposition, but they may be more visible.

The accuracy of the total force depends only weakly on choice of the grid size $K$, and so in practice we choose $K$ to optimize run time. The far-field work in the convolution scales with the number of grid cells, while the near-field work scales (for a fixed total number of particles) roughly as the number of particles per cell, which is inversely with the number of grid cells.  Optimal values are typically 30--100 particles per cell, depending on the relative speed of the GPU and CPU.

\subsubsection{Computation of the Cell Multipoles}\label{sec:multipoles_simd}
The far-field force is generated by computing multipoles in every cell, performing the 3D discrete Fourier transform on the grid of multipoles, convolving with the derivatives tensor, applying the inverse Fourier transform, and evaluating the force from the resulting Taylor series of the potential in every cell.

The first and last steps---the cell-wise multipoles and Taylors---are the only steps that ``see'' the particles.  Therefore, these steps are key to optimize, since their work scales as $\mathcal{O}(p^3N)$.

It is computationally convenient to compute the multipoles in the \textit{complete} basis, consisting of $(p+3)(p+2)(p+1)/6$ values given by the usual Cartesian multipoles for a restricted set of indices \citep{Metchnik_2009}.  For a set of $n$ point masses with displacements $(x_q,y_q,z_q)$ from the origin of cell $(i,j,k)$, the multipoles can be written as:
\begin{align}\label{eqn:multipoles}
    M_{i,j,k}^{a,b,c} = \sum_q^n x_q^a y_q^b z_q^c, && \text{for}
    \begin{cases}
    0 \le a \le p \\ 
    0\le b \le p-a \\
    0\le c \le p-a-b
    \end{cases}
\end{align}
After evaluation in this basis, the multipoles are converted at small computational cost to a \textit{reduced} basis of $(p+1)^2$ values, exploiting the trace-free property of the multipoles to losslessly encode the gravitational potential \citep{Hinsen_Felderhof_1992,Metchnik_2009}.  The subsequent Fourier transforms and disk storage operate in the reduced basis.  For $p=8$, the reduced basis has about half the terms of the complete basis.

We now turn to efficient computation of Eq.~\ref{eqn:multipoles}.  In C-like pseudo-code, this equation may be written as:

\begin{minipage}[]{\linewidth}
\begin{minted}{C}
int cml = (order+3)*(order+2)*(order+1)/6;
double M[cml] = {0};
for(int q=0; q<n; q++){
    double mx = 1.;
    for(int a=0; a<=order; a++){
        double mxy = mx;
        for(int b=0; b<=order-a; b++){
            double mxyz = mxy;
            for(int c=0; c<=order-a-b; c++){
                M[a,b,c] += mxyz;
                mxyz *= Z[q];
            }
            mxy *= Y[q];
        }
        mx *= X[q];
    }
}
\end{minted}
\end{minipage}

While succinct, this quadruply-nested loop does not compile to efficient code.  In particular, it is difficult for the compiler to unroll the inner triple loop over multipole orders, so the resulting program flow has many tests and jumps.  However, since the order is usually fixed for a given simulation, it is straightforward to write a meta-code that manually unrolls the inner triple loop.  This results in a substantial speedup.

\begin{figure}
    \centering
    \includegraphics{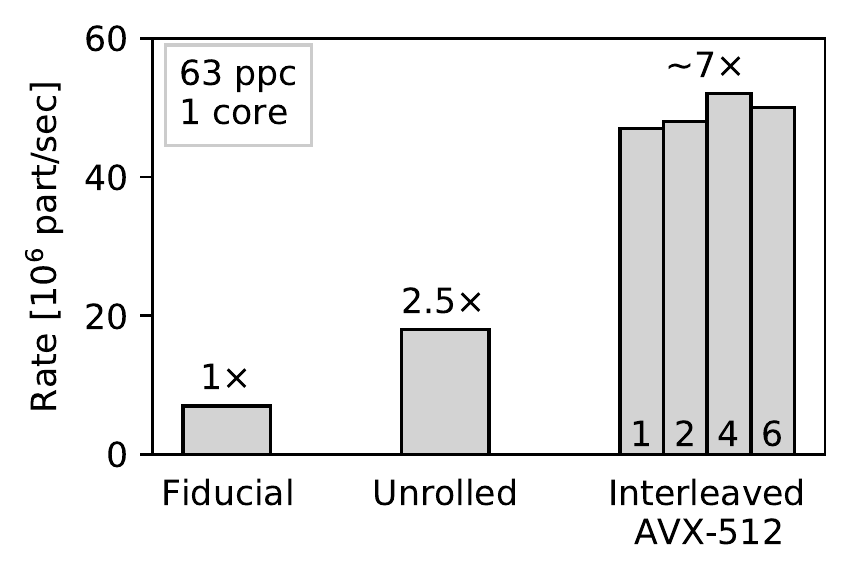}
    \caption{The effect of successive optimizations on the multipoles computation.  The \textit{fiducial} implementation has the triple-nested loop over multipole order; the \textit{unrolled} implementation unrolls that. The \textit{interleaved} version uses 1, 2, 4, or 6 AVX-512 vectors.  All cells have 63 particles to trigger the ``remainder'' iterations.}
    \label{fig:multipoles_optimization}
\end{figure}

Furthermore, modern processors contain single-instruction, multiple-data (SIMD) instructions, also known as vector instructions, to allow simultaneous processing of multiple values, typically 256 or 512 bits wide (4 or 8 doubles).  These can substantially improve the throughput of a FLOP-bound code; that is, a code where the floating-point operations dominate the runtime, rather than memory movement, CPU pipeline dependencies, or other resource constraints.  Tallying the arithmetic operations in the above pseudo-code, we may naively expect 384 FLOP per 24 byte particle, which, while only a rough estimate, indicates that this code may be in the FLOP-bound regime, with 16 FLOP/byte.

It is therefore important to ensure the compiler emits SIMD optimizations for the code.  The multipoles meta-code uses intrinsics (built-in functions that map directly to an assembly instruction) for maximum control, but in other areas of the code, we use less invasive approaches.  This includes source-level compiler directives (pragmas), and mild manipulation of the source code guided by compiler diagnostics or assembly output.

Many processors also contain SIMD ``fused multiply-add'' (FMA) operations to evaluate expressions of the form \ttt{a*b + c} in a single instruction, where each operand is allowed to be a SIMD vector.  This is a potential factor of 2 increase in floating-point throughput, compounded with the SIMD speedup.

However, in the above multipoles pseudo-code, none of the computations are expressed as FMA.  But by precomputing $p$ powers of $z^c$ in a new array \ttt{Zc}, we can achieve such a transformation, effectively replacing the body of the innermost loop with the following FMA pattern:
\begin{minted}{C}
        M[a,b,c] += mxy*Zc[c];
\end{minted}

On some platforms, we have found such a transformation beneficial, but on others, the performance is the same or worse. This may be due to vector register or memory operand pressure: a new \ttt{Zc[c]} vector is required for each loop iteration.



Even with properly vectorized SIMD FMA, we will not exploit the full floating-point capability of the processor.  Floating-point instructions tend to have latencies of several cycles between the time they are issued and the time they retire (return their results).  New instructions may be issued while waiting for results, but only if they do not depend on the results of previous instructions, which the multiply and accumulate instructions do (at least as phrased in our pseudo-code).

Therefore, we interleave the entire multipole computation for multiple vectors of particles.  This is a trade-off between masking instruction latency and register pressure, but is especially beneficial on processors that can launch multiple vector instructions per cycle.

The SIMD and vector-level interleaving are very efficient at dealing with dense cells, and in particular those whose occupation number is a multiple of the SIMD vector width times the number of interleaved vectors.  But an exact multiple is rare, so it is important to efficiently deal with ``remainder'' particles.  Indeed, in clustered cosmological simulations where most cells have low occupation number (due to the clustered distribution of matter), the typical cell may have a particle count lower than the minimum interleaved SIMD width (even if the typical particle exists in a dense cell).

Our approach to remainder particles is to generate ``step-down'' iterations of the interleaving.  The meta-code first emits a fully interleaved SIMD implementation that will process as many particles as it can, until fewer particles remain than the interleaved-SIMD width.  Then, the meta-code emits step-down iterations, with one fewer interleaved vector at each step, until only one vector remains.  Only one of these step-down iterations will be needed per cell (the largest one that is narrower than the remaining particle count).  Then, if any particles remain, they must be narrower than the width of a single SIMD vector.  These are either treated with masked vector operations on platforms that support it, or a scalar loop on platforms that do not.


Figure \ref{fig:multipoles_optimization} shows the effect of several of these optimizations on a single core of an Intel Xeon processor with two AVX-512 FMA units.  63 particles was used to trigger the fully interleaved loop, the step-down iteration, and the remainder iteration.  The FMA version of the multipoles computation was used.  At a peak of 52 million particles per second, this is about 20 GFLOPs, or 35\% of peak single-core performance.  The absolute performance is modestly smaller in real, multi-threaded contexts, but the relative performance of the successive optimizations remains similar in our tests.  All three give the same answer to within $10^{-15}$.

The Taylors computation is somewhat symmetric to the multipoles computation, except that the output is a 3-vector of acceleration for every particle.  There is correspondingly about 3 times more floating-point work.  There are also several options regarding precomputation, in the vein of the $z^c$ computation for the multipoles.  The spirit of these optimizations (meta-code, unrolling, SIMD, FMA, intrinsics, interleaving, remainders) is similar to what we have already discussed.

\subsection{GPU Near-Field Computation}\label{sec:gpu_data_model}
\subsubsection{Overview}
We now turn to the near-field force computation.
This consists of every cell interacting with itself and its 124 nearest neighbor cells (for near-field radius $R=2$) using open boundary conditions and Newtonian gravity, or a softened form thereof.  We evaluate forces on one slab at a time, but must have 5 in memory: the central slab is the ``sink'' slab that receives forces from itself and its neighbor slabs, and the neighbor slabs are ``source'' slabs.  Our approach is to compute all $N^2$ pairwise interactions, leveraging the massive parallelism of GPUs to accelerate the computation.  In this model, the efficiency of the simple $N^2$ approach outweighs the gains of a more complicated (e.g.~tree) approach for all but the most clustered problems.  The architecture of the GPU demands some care in arranging the particle data and interaction sets for efficient computation; we discuss this data model here.  We will use the case $R=2$ for concreteness, so all instances of ``5 cells'' or ``5 slabs'' may be replaced by ``$2R+1$'' cells or slabs.

\begin{figure}
    \centering
    \includegraphics[width=\columnwidth]{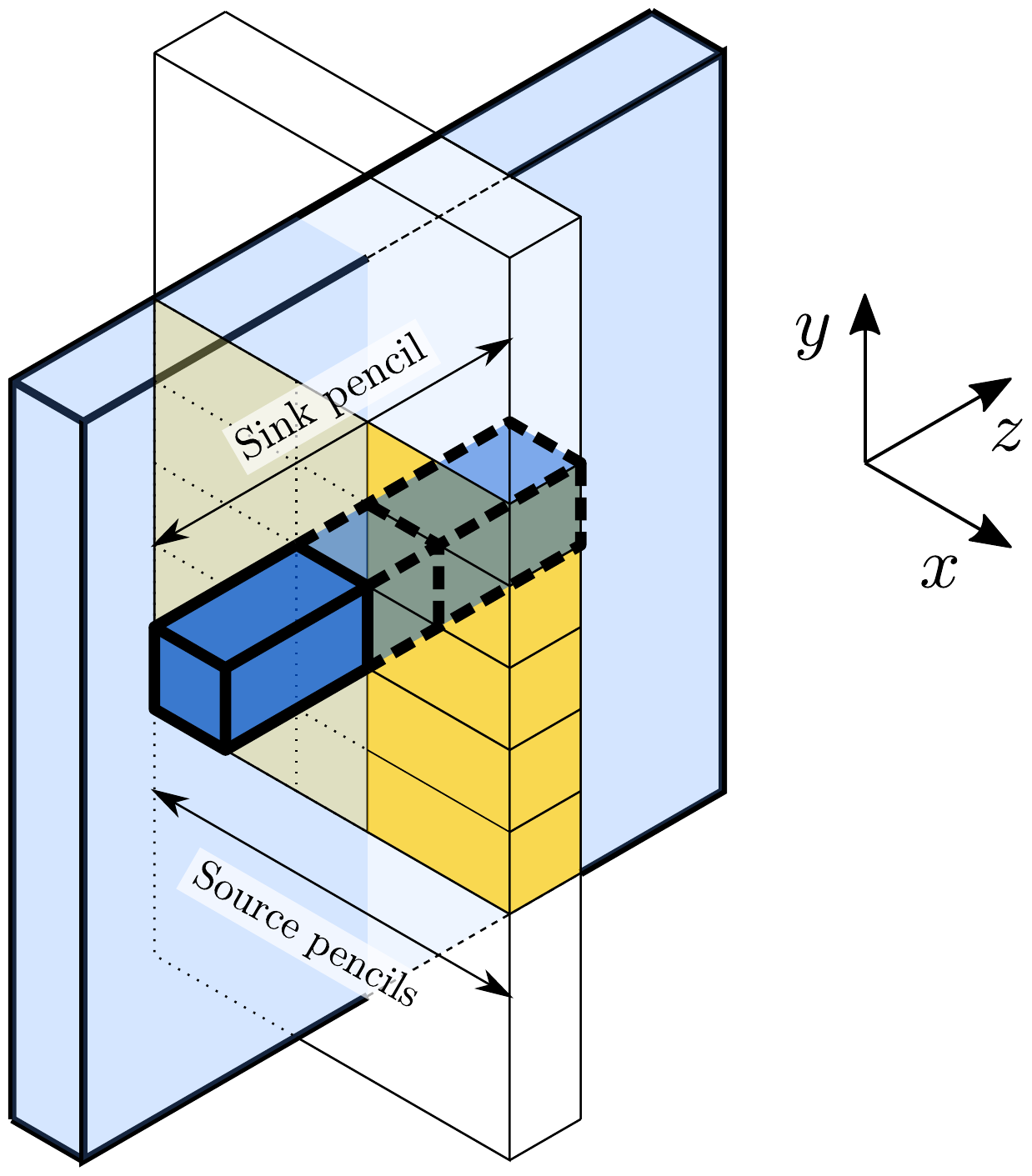}
    \caption{The pencil-on-pencil GPU data model.  All particles in the sink pencil (blue) receive accelerations from all particles in the 5 source pencils (yellow).  The sink pencil resides in a single slab, while the source pencils cross slabs.}
    \label{fig:pencil_on_pencil}
\end{figure}

\subsubsection{Pencil-on-Pencil}
We arrange the computation as \textit{pencil-on-pencil} interactions, where every pencil is a linear block of 5 cells.  Sink pencils run in the $z$ direction, while source pencils run in the $x$ direction (across slabs).  A sink pencil centered at cell $(i,j,k)$ will interact with the 5 source pencils centered at $(i,j+b,k)$ for $b\in [-2,2]$ (Figure \ref{fig:pencil_on_pencil}).  Geometrically, one can think of each sink pencil being acted on by the plane of 5 source pencils that intersects its center perpendicularly.  By having each sink cell appear in 5 sink pencils, centered at $(i,j,k+c)$ for $c\in [-2,2]$, we thereby include every needed pairwise interaction exactly once.  Each cell accumulates 5 partial accelerations, one for each of its parent pencils.

All of the sink pencils for a given $i$ and range of $j$ are indexed in a \ttt{SetInteractionCollection} (\ttt{SIC}) object along with the corresponding source pencils.  Upon pencil construction, particle positions are adjusted from cell-centered to pencil-centered: this allows the GPU to remain agnostic to cell divisions and seamlessly handles the periodic wrap.  Sink particles are arranged into blocks with a \texttt{BlockSize} of 64 particles (or some other multiple of the atomic CUDA warp size of 32 threads).  The membership of blocks in pencils defines which sink blocks must interact with which source blocks---this is how the GPU views the pencil-on-pencil model: interactions of blocks rather than pencils.

Pencils are essentially virtual indexing structures until the particles are loaded for staging to the GPU.  At that point, all pencils are explicitly constructed (and thus 5 copies of every sink particle are made).  Each copy is offset to pencil coordinates as discussed above; thus, the sink and source coordinate systems differ by at most a $y$ offset.  The $y$ offset is stored for every pencil interaction and is also passed to the GPU, where it is applied on the fly.

One CUDA kernel launch is executed for each \ttt{SIC}.  Each CUDA kernel launches with a grid of \texttt{NSinkBlocks} thread blocks, each containing \texttt{BlockSize} threads.  On the GPU, each thread is responsible for one sink (this is why the particle block size must be a multiple of the thread warp size).  Each thread loads its sink into a register, and then the work loop begins: each thread loads one source into shared memory and waits at a barrier, such that all threads have access to \texttt{BlockSize} sources at a time.  Then each thread loops over sources and computes the $1/r^2$ interaction, or a softened form thereof.  Accelerations are accumulated into per-sink registers, which are saved into global GPU memory at the end of the kernel.

Why go through all this trouble to construct pencils?  The simplest data model would be to compute cell-on-cell interactions, but that would be substantially less efficient.  An NVIDIA V100 GPU has 900 GB/s of GPU memory bandwidth, or 75 Gsources/s.  But the compute rate is 15 TFLOPS, or $\sim 650$ Gforces/s.  Thus, each source should be used at least 10 times per load to avoid being bandwidth-limited.  Packing the sinks into pencils that fill thread blocks ensures 64 uses per load.  Furthermore, NVLink can only transfer data from host memory to the GPU at 35 GB/s, measured on Summit using NVIDIA's bandwidth tool, so we would like to use each source at least 250 times per transfer.  With each source acting on at least 5 sink pencils (possibly more across different $j$), this means we only need 10 particles per cell which is achievable outside of sparse voids.

\subsubsection{Host-Device Communication}

The \ttt{SetInteractionCollection} construction happens as soon as the positions and cell info for 5 slabs are loaded.  No particle copies happen at this time; the \ttt{SIC} instead constructs the necessary indexing information for a later copy.  The \ttt{SIC} is pushed to a work queue corresponding to its NUMA node that is monitored by several CPU threads; when a thread is free, it pops a \ttt{SIC} from the queue and begins executing it.  First, the thread constructs pencils by copying particles from the slabs to pinned memory, applying coordinate offsets on-the-fly.  Then, it launches the CUDA copies, the main work kernel, and the acceleration copy-back.  The thread then blocks while waiting for the results.  Finally, once all the accelerations are back in host memory, the 5 partial accelerations are combined into one total acceleration; this reduction is performed directly into the acceleration slab.  The result is the final near-field force for every particle that was part of the \ttt{SIC}.

We use CUDA ``pinned memory'' as the staging area where we construct pencils to send to the GPU and receive accelerations back from the GPU.  Pinning memory locks RAM pages such that they have a guaranteed physical address (not just a virtual address).  This enables direct memory access (DMA) transfers between host RAM and GPU memory with no CPU intervention.

The copy of particles into pinned memory does apply extra memory pressure, but it ensures optimal data transfer rate to the GPU and allows the CUDA stream to proceed unblocked.  The initial pinning of memory is slow, however---at least a few seconds, depending on the amount of pinned memory.  This is a noticeable overhead in the \Abacus model where a new process is invoked for every time step.  To mask this latency, we overlap convolution work with the GPU startup, which is effective until late times when the GPU work becomes the rate-limiting factor.


We typically employ three CPU threads per GPU.  Each thread manages one CUDA stream and has its own pinned memory buffers.  The thread executes the pencil construction and acceleration co-addition (essentially all pinned memory work).  Using two to three streams per GPU ensures overlap of host-device transfers and GPU compute.


\subsubsection{Force Kernel}\label{sec:gpu_kernel}

This careful data packaging would be for naught if our force kernel were slow to compute.  We adopt the spline softening force kernel, introduced in Section \ref{sec:softening}, which is both amenable to GPU execution and explicitly switches to $1/r^2$ at finite radius.  From a computational perspective, the key aspects are the small number of floating-point operations (FLOPs) per interaction (about 26, including the reciprocal square root) and the implementation of the hand-off to the $1/r^2$ form with a \texttt{min} instead of a conditional for protection against costly conditional jumps.


The force kernel also computes a estimate of the local density, exploiting the fact that a top-hat density estimate simply requires incrementing a counter if a particle is within radius $b$.  The GPU is already computing distances to nearby particles for the gravitational computation, so this density estimate comes nearly for free.  In detail, we find it useful to accumulate an apodized top-hat density estimate $1-r^2/b^2$ (for $r<b$), which is computed as $b^2-r^2$ on the GPU---a single subtraction.  By promoting 3-vectors of accelerations to 4-vectors, the density estimate may be stored as the 4th component of acceleration.  We find the density computation and extra acceleration storage to be small overheads.

We use the NVIDIA's CUDA programming language for our GPU implementation, as it is the most mature GPU programming language, and the NVIDIA GPUs have wide adoption in the supercomputing community.  A port to another architecture, such as AMD Radeon Instinct or Intel Xe, should be straightforward, as the basic way in which the data model exposes massive parallelism is portable.

\subsection{Force Accuracy}\label{sec:force_accuracy}
\subsubsection{Overview}
We consider three direct tests of \Abacus's force accuracy: a comparison of a random particle distribution against an Ewald summation; an examination of force residuals on a lattice; and a comparison of a single-precision, multipole order 8, \LCDM simulation against the same snapshot evaluated in double precision at multipole order 16.  These three tests are presented in the following sections.

Before examining these tests, we note one aspect of the code that aids the accuracy, which is the representation of particle positions as offsets relative to cell centers. These offsets, and the velocities, are stored in single precision (32-bit floats), but with typical values of a few hundred to a few thousand cells per dimension, the positions retain an additional 9--10 bits of mantissa beyond the nominal 24 in the IEEE-754 floating-point representation.  Multipole and Taylor data is stored on disk as 32-bit floats, but all internal far-field computations (Multipoles, Taylors, and FFTs) are performed in double precision to avoid potential buildup of round-off error.

In addition to the tests presented in this section, \cite{Garrison+2016,Garrison+2018,Garrison+2019} contain published results from \Abacus, including validation tests against analytic theory, against other codes and emulators, and in internal convergence tests.

\subsubsection{Ewald}\label{sec:ewald}
In the Ewald test, we use a random distribution of 65536 particles whose periodic forces were computed with brute-force Ewald summation \citep{Ewald_1921} in quad-double (256-bit) precision (Marc Metchnik, private communication).  This used a code separate from any \Abacus infrastructure and therefore serve as an external set of reference forces.  Here, we compare these forces relative to \Abacus forces using \ttt{CPD} 11, or 49 particles per cell, and multipole order 8 (both typical values in our \LCDM simulations).

\begin{figure}
    \centering
    \includegraphics{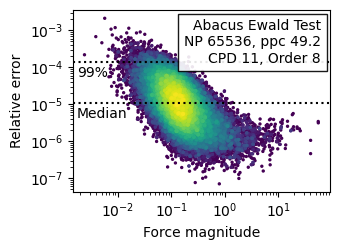}
    \caption{Fractional Ewald test force error versus reference force magnitude.  The reference is the brute-force solution from Ewald summation in quad-double precision.  The median is $1.1\times 10^{-5}$ and the 99th percentile is $1.4\times 10^{-4}$.}
    \label{fig:ewald}
\end{figure}

The results are shown in Figure \ref{fig:ewald}.  We find that Abacus's 99\% and median fractional errors are $1.6\times10^{-4}$ and $1.2\times10^{-5}$, respectively.  The absolute forces span 4 orders of magnitude, with higher fractional errors appearing on particles with low force magnitude (smaller denominator).

\begin{figure}
    \centering
    \includegraphics[width=\columnwidth]{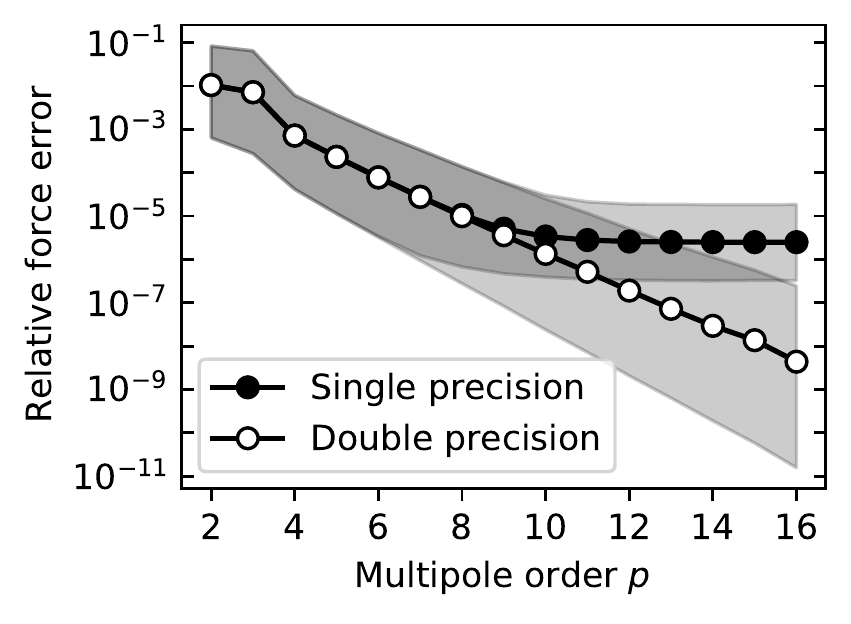}
    \caption{Fractional Ewald test force error versus multipole order in single and double precision.  The test is the same as in Fig.~\ref{fig:ewald}, but showing the effect of multipole order.  The shaded bands indicate the 1st percentile to 99th percentile, with the circles marking the median.  There is little gain in accuracy past multipole order 10 in single precision (at least in the random particle distribution employed in this test).}
    \label{fig:ewald_allorders}
\end{figure}

Figures \ref{fig:ewald_allorders} show variations in multipole order for single and double precision.  In single precision, gains in accuracy diminish beyond order 10.  In double precision, we see no floor in the accuracy to multipole order 16, the highest order tested.


\subsubsection{Homogeneous Lattice}\label{sec:lattice}

\begin{figure}
    \centering
    \includegraphics[width=\columnwidth]{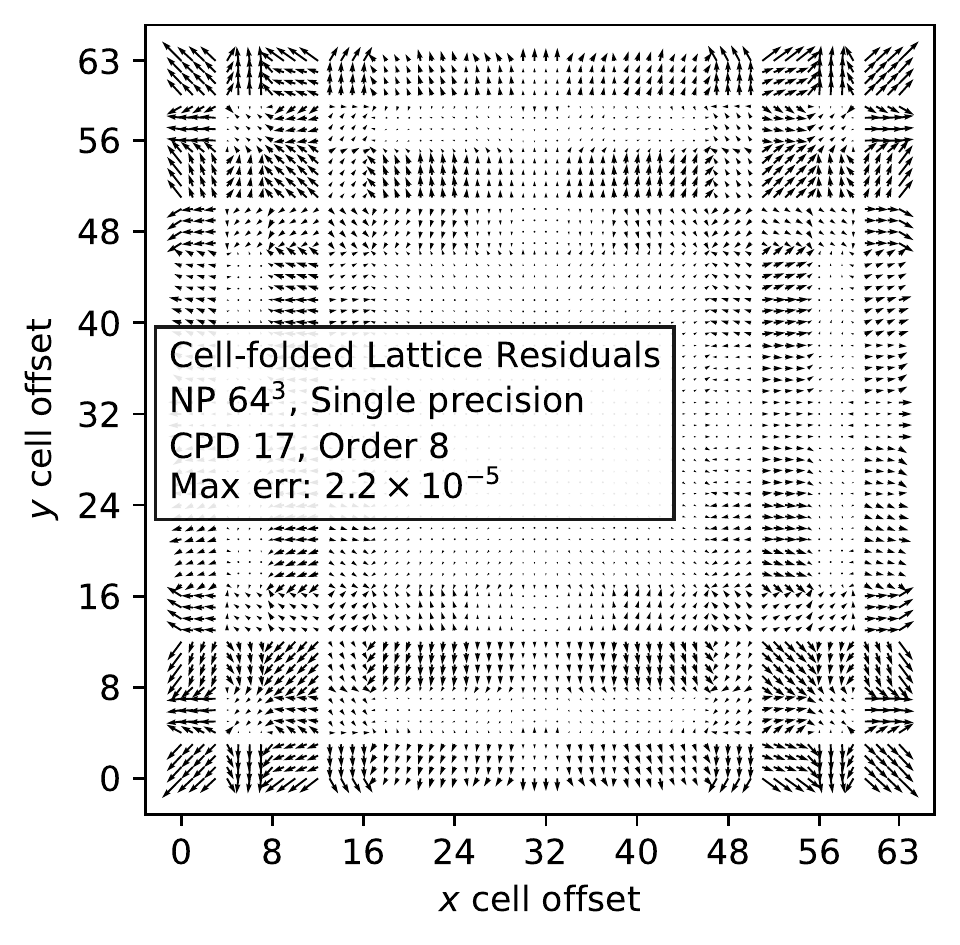}
    \caption{Residual forces on a pure cubic lattice, expressed as an equivalent displacement.  By choosing a cells-per-dimension value that is co-prime with the particles-per-dimension, each particle samples a different cell location.  This figure is a map of those locations (i.e.~a map of a single cell) for the $z=0$ particle plane (a cell edge).  The maximum error is $2.3 \times 10^{-5}$ in units of equivalent interparticle spacing (see text).  The ``banding'' structure arises from the aliasing of the particle lattice against the cell grid.}
    \label{fig:lattice_residuals}
\end{figure}

In the homogeneous lattice test, we establish a uniform grid of particles such that the forces should be zero by construction.  We choose a number of particles-per-dimension that is co-prime with \ttt{CPD} so that every particle samples a unique cell location.  For $p=8$, $N=512^3$, and \ttt{CPD} 125, the maximum deviation is $2.6\times10^{-5}$, in units of the displacement that would produce that force under the Zel'dovich Approximation \citep{Zeldovich_1970}, expressed as a fraction of the inter-particle spacing.

\begin{figure}
    \centering
    \includegraphics[width=\columnwidth]{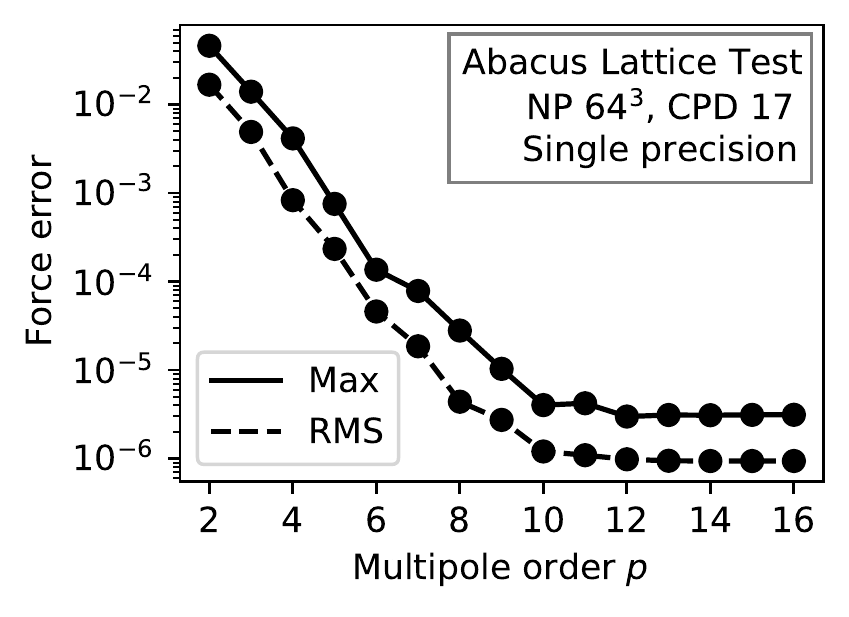}
    \caption{Lattice error on the 3D force magnitude (maximum, solid line; root mean square, dashed line) versus multipole order.  Units are same as in Fig.~\ref{fig:lattice_residuals}.}
    \label{fig:lattice_all_orders}
\end{figure}

The results are shown as a function of cell position in Figure \ref{fig:lattice_residuals}.  Variation of multipole order is shown in Figure \ref{fig:lattice_all_orders}.  The RMS force error is about $4\times10^{-6}$ at order 8, with diminishing returns past order 10 due to single precision, as with the Ewald test.  The units of the RMS are again equivalent fractional displacements.

The banding structure of the force residuals in Figure \ref{fig:lattice_residuals} arises from the ``beats'' of the particle lattice against the cell grid.  In particular, with 64 particles-per-dimension and 17 cells-per-dimension, each cell must either have 3 or 4 particles.  Transitioning between cells results in a discontinuous jump in the pattern of particles in the near-field width of 5 cells: 3-4-4-4-3, or 4-4-4-3-4, etc.

The homogeneous lattice configuration is classically a difficult one for $N$-body simulations, as the density perturbations are small and the uniformity means that particles in shells of all radii contribute equally to the force.  There is no clustering for tree codes to exploit.  Some codes apply stricter force error tolerances at early times to combat this difficulty (e.g.~higher multipole order at high redshift in \ttt{PKDgrav}, \citealt{Potter+2017}).

\begin{figure}
    \centering
    \includegraphics[width=\columnwidth]{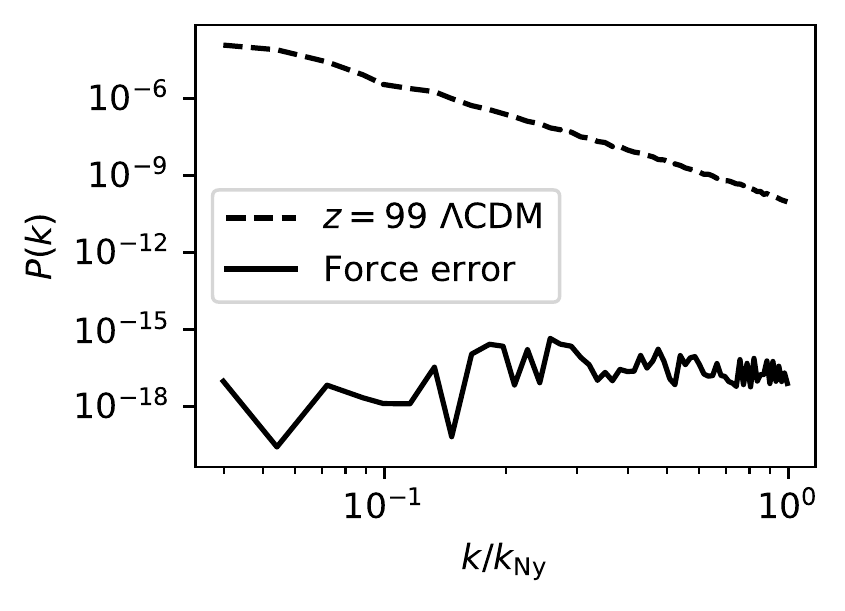}
    \caption{The power spectrum of the $p=8$ lattice force errors (Fig.~\ref{fig:lattice_residuals}) compared to a typical $z=99$ power spectrum for Zel'dovich Approximation initial conditions in a $\Lambda$CDM cosmology (for particle mass $2\times 10^9 \hMsun$, or top-hat density RMS $\sigma=0.06$ at the interparticle spacing).  The power spectrum is evaluated on one component of the (equivalent) Lagrangian displacements, in units of the interparticle spacing.}
    \label{fig:force_err_pk}
\end{figure}

\Abacus does well in this configuration.  The errors are equivalent to a tiny fraction of the interparticle spacing; the power in the error is at least 6 orders of magnitude below the typical power in the initial conditions (Figure \ref{fig:force_err_pk}).  This high accuracy supports our derivation of 2LPT in configuration space from direct force evaluations \citep{Garrison+2016}.

The homogeneous lattice test is also useful because it readily scales to massive problems, since the lattice can be generated on-the-fly.  This is useful for testing the parallel force solver.

\subsubsection{\LCDM}\label{sec:lcdm_order16}
Neither the Ewald nor homogeneous lattice tests are representative of the clustered particle distributions observed in late-time cosmological simulations.  While our force solver is largely agnostic to particle clustering, it is nonetheless prudent to test the forces with realistic configurations.

We take a snapshot of a $z=0$ LCDM simulation with $512^3$ particles in a 150 \hMpc box and compare the forces to the same snapshot converted to double precision and evaluated with multipole order 16.  While not an absolute benchmark, the force on most particles should be converged to single-precision accuracy (e.g.~Figure \ref{fig:ewald_allorders}).

\begin{figure}
    \centering
    \includegraphics[width=\columnwidth]{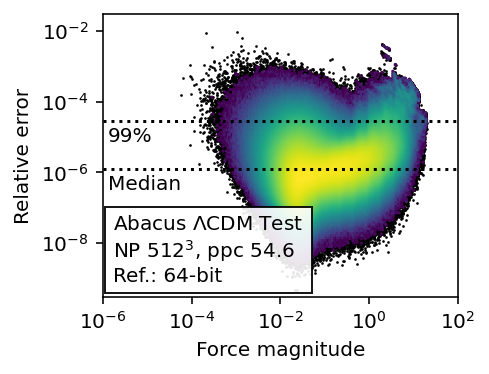}
    \caption{Fractional force errors in a $512^3$ \LCDM simulation ($z=0$, particle mass $2\times 10^{9}\ \hMsun$).  The forces are from the standard \Abacus configuration of single precision, multipole order 8, compared to a reference solution of double precision, multipole order 16. The median fraction error is $1.2\times 10^{-6}$, and the 99\% is $2.9 \times 10^{-5}$.}
    \label{fig:lcdm_force_err}
\end{figure}

\begin{figure}
    \centering
    \includegraphics[width=\columnwidth]{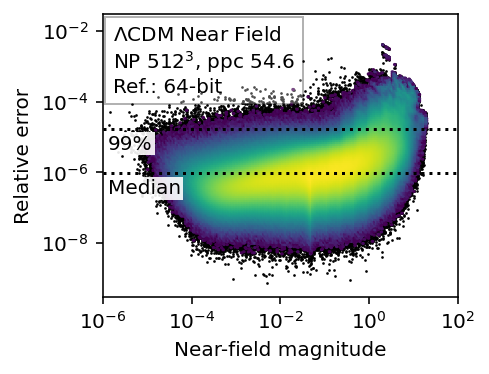}
    \caption{Near-field force errors for the \LCDM simulation of Fig.~\ref{fig:lcdm_force_err}.  The median error is $9.2\times 10^{-7}$; the 99th percentile is $1.7\times 10^{-5}$.  The increase at high force magnitude is due to loss of precision in the accumulation of accelerations in single precision.}
    \label{fig:lcdm_near_err}
\end{figure}

\begin{figure}
    \centering
    \includegraphics[width=\columnwidth]{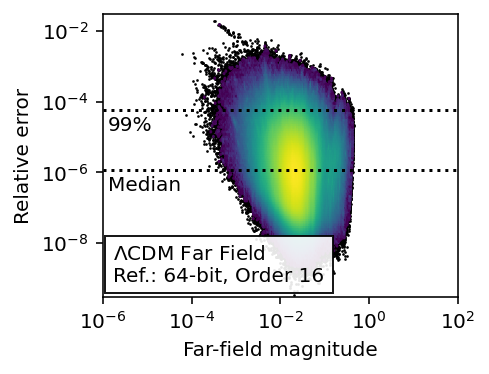}
    \caption{Far-field force errors for the \LCDM simulation of Fig.~\ref{fig:lcdm_force_err}.  The median is $1.1 \times 10^{-6}$; the 99th percentile is $5.6\times 10^{-5}$.}
    \label{fig:lcdm_far_err}
\end{figure}

The results are shown in Figure \ref{fig:lcdm_force_err}, and Figures \ref{fig:lcdm_near_err} \& \ref{fig:lcdm_far_err} for the near- and far-field shown separately.  The median force error is $1.2\times10^{-6}$, which is better than the previous tests likely due to the increased amplitude of the near field, which is exact save for computational roundoff error.  Indeed, we see higher force accuracy in the near-field figure than the far-field, as expected.  We can also see that the bimodality in the total force error arises from the different error properties of the near and far field.

\begin{figure*}
    \centering
    \includegraphics[width=\textwidth]{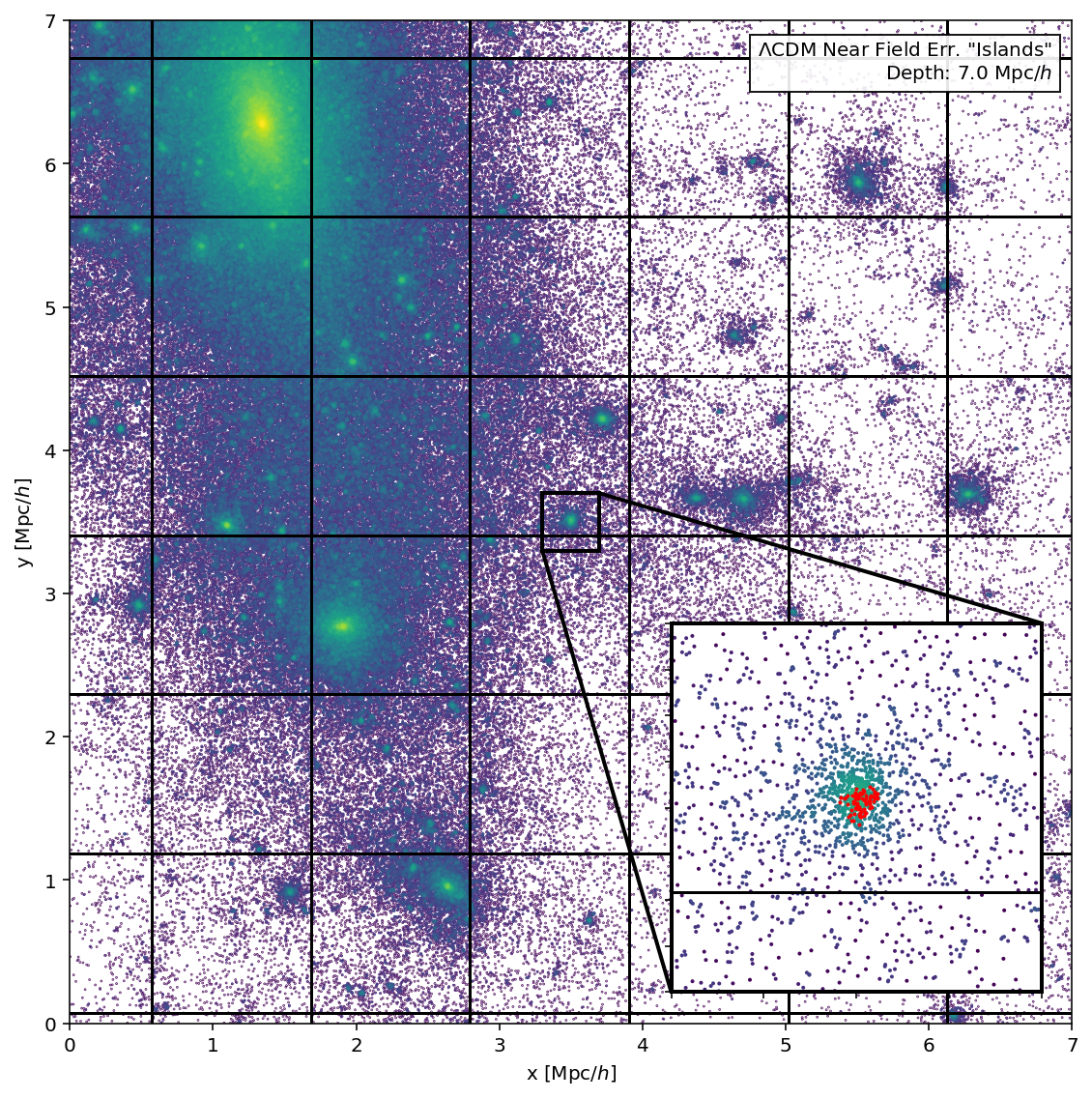}
    \caption{Location of the outliers in the near-field force error.  In Figure \ref{fig:lcdm_near_err}, two small ``islands'' of high fractional force error are seen (about 0.1\%); these particles are plotted in red in the inset.  The large halo in the corner of the near field induces a loss of precision the accumulation of the acceleration.}
    \label{fig:near_islands}
\end{figure*}

The near-field force error increases towards larger force magnitude in Figure \ref{fig:lcdm_force_err}, which is a symptom of loss of precision in the accumulation of accelerations onto a large accumulator.  This could be mitigated in future versions of \Abacus with Kahan summation \citep{Kahan_1965}, intermediate accumulators, or higher-precision global accumulators.  The overall degradation of the force precision is still mild, except in a few extreme cases, seen as two ``islands'', or stripes, of high force error (0.1\%).  The location of these particles is plotted in Figure \ref{fig:near_islands}, where they are all seen to reside in a small halo with a large halo in the corner of the near-field.  The large sum of accelerations due to that halo likely makes it difficult to retain precision from additional small-magnitude accelerations.

\begin{figure*}
    \centering
    \includegraphics[width=\textwidth]{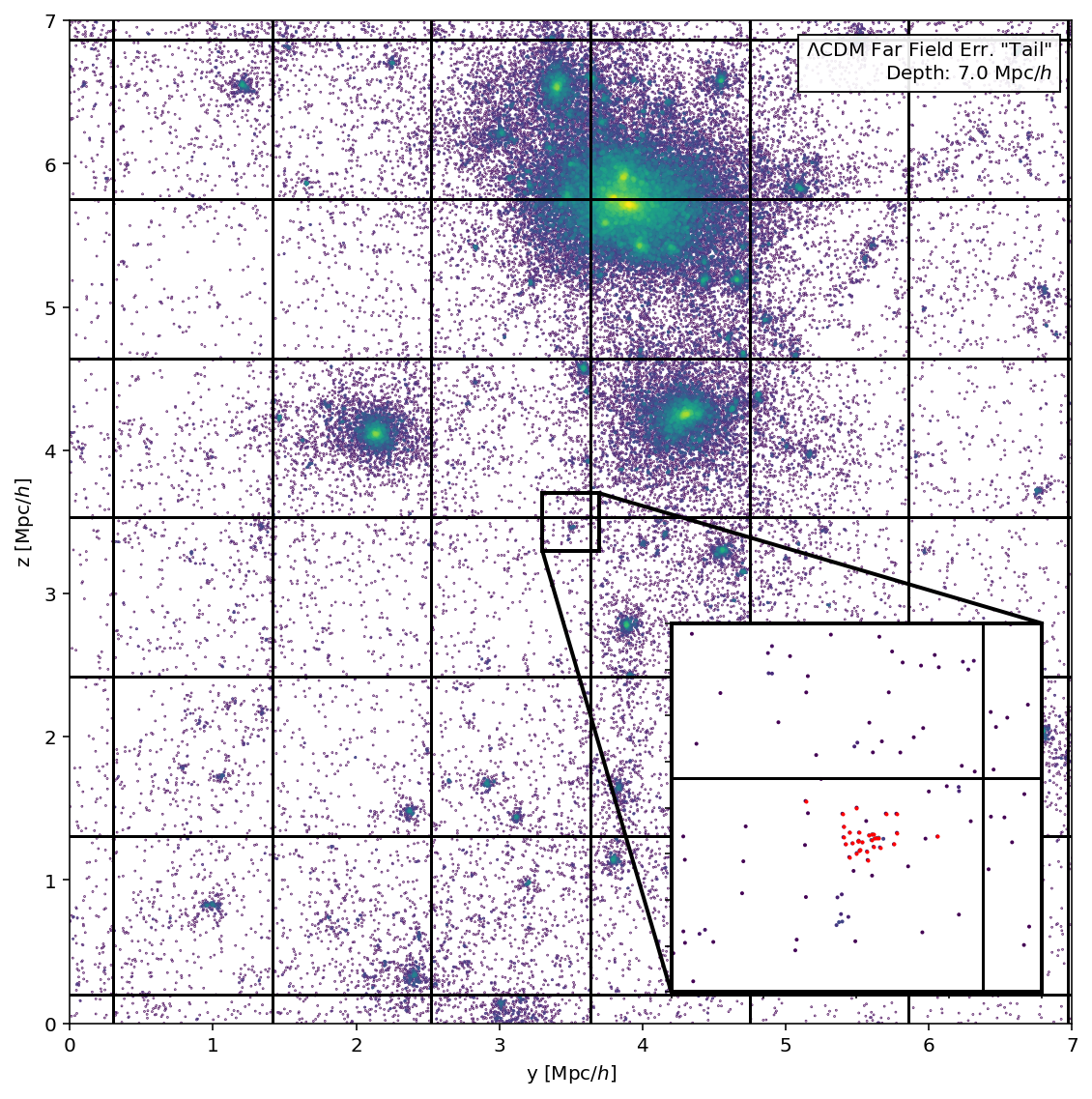}
    \caption{Location of the far-field force error ``tail''.  In Figure \ref{fig:lcdm_far_err}, a small ``spur'' of high fractional force error particles is seen (about 1\%); these particles are plotted in red in the inset.  The error likely arises from splitting of a large halo across the near-field/far-field boundary, although this 1\% far-field error is greatly suppressed by a larger, more accurate near-field.}
    \label{fig:far_tail}
\end{figure*}

A tail of large (1\%) far-field error can be seen Figure \ref{fig:lcdm_far_err} (showing just the far-field force).  This appears to occur when a halo is split by the near-field/far-field transition (Figure \ref{fig:far_tail}, where the tail is marked in red).  But no such 1\% error is seen in the total force, because the near-field force tends to be both larger and more accurate.



\section{Event-Driven Slab Pipeline}\label{sec:pipeline}
\subsection{Pipeline}
\begin{figure}
    \centering
    \includegraphics[width=\columnwidth]{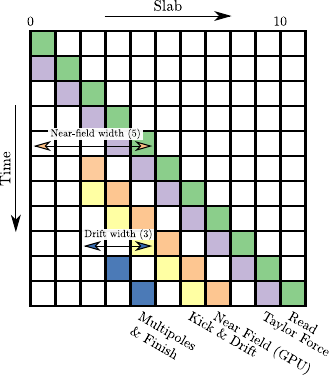}
    \caption{A schematic illustration of the \Abacus slab pipeline.  A rolling window of slabs is loaded into memory, starting with the Read action (green), and ending with the Finish action (blue).  Each colored box indicates an action currently in progress, which may run concurrently with other actions in the same row.  The two diagonal ``gaps'' are due to cross-slab dependencies: the near-field computation requires $\pm2$ slabs in memory, and the Multipoles require that both neighboring slabs have Drifted so that incoming particles may be received. The slab pipeline is complete when the Finish action has executed on all slabs.  See Table~\ref{tbl:pipeline} for action descriptions.
    }
    \label{fig:pipeline_grid}
\end{figure}

The \Abacus particle update cycle---load, compute forces, kick, drift, store---is expressed as a \textit{slab pipeline}.  A slab is a plane of $1\times K\times K$ cells (one cell wide in the $x$ direction).  The pipeline is implemented as an event loop with a set of \ttt{Dependency} objects, each with \textit{preconditions} and an \textit{action}.  The preconditions express dependencies between pipeline actions; a given action can only execute after its preconditions are fulfilled.  For example, \texttt{Kick[2]}, the velocity update for slab 2, can only execute once \texttt{TaylorForce[2]} and the asynchronous GPU work (as launched by \ttt{NearForce[2]}) have completed. Dependencies may also cross slabs: \texttt{NearForce[2]} must wait for \texttt{LoadSlab[0]} through \texttt{LoadSlab[4]} (the near-field radius must be in memory). Figure~\ref{fig:pipeline_grid} illustrates the slab pipeline, and Table~\ref{tbl:pipeline} shows a selection of pipeline stages, preconditions, and actions.

\fboxsep=1mm \fboxrule=0.5mm
\definecolor{C1}{HTML}{7fc97f}
\definecolor{C2}{HTML}{beaed4}
\definecolor{C3}{HTML}{fdc086}
\definecolor{C4}{HTML}{ffff99}
\definecolor{C5}{HTML}{386cb0}
\begin{deluxetable*}{lll}
\tablecaption{A selection of \Abacus pipeline stages}
\tablehead{\colhead{Action} & \colhead{Precondition} & \colhead{Description}}
\startdata
\raisebox{0.7ex}{\fcolorbox{black}{C1}{\null}}
\ttt{Read} & \ttt{Kick[N-3]} & Load slabs into memory, reading ahead of the \ttt{Kick} a small amount \\
\raisebox{0.7ex}{\fcolorbox{black}{C2}{\null}}
\ttt{TaylorForce} & \ttt{Read[N]} & Compute far-field forces using the Taylor expansion of the potential \\
\raisebox{0.7ex}{\fcolorbox{black}{C3}{\null}}
\ttt{NearForce} & \ttt{Read[N-R,N+R]} & Launch near-field forces on the GPU \\
\raisebox{0.7ex}{\fcolorbox{black}{C4}{\null}}
\ttt{Kick} & GPU done [N] \& \ttt{TaylorForce[N]} & Update velocity from $t-1/2$ to $t + 1/2$ \\
\raisebox{0.7ex}{\fcolorbox{black}{C4}{\null}}
\ttt{Drift} & \ttt{Kick[N]} & Update position from $t$ to $t + 1$ \\
\raisebox{0.7ex}{\fcolorbox{black}{C5}{\null}}
\ttt{Finish} & \ttt{Drift[N-1,N+1]} & Build merged slabs, compute multipoles, and write results\\
\enddata
\tablecomments{The current slab, \ttt{N}, is different for each pipeline stage. \ttt{R} is the near-field radius, typically 2. Colors correspond to Fig.~\ref{fig:pipeline_grid}.}
\label{tbl:pipeline}
\end{deluxetable*}


This event-driven model makes it easy to incorporate asynchronous events, such as completion of I/O, GPU kernel execution, or MPI communication.  With a properly expressed set of dependencies, the maximum amount of work possible can execute while waiting for external events.  Each dependency action will run exactly \ttt{CPD} times, once per slab, and the pipeline exits when the \texttt{Finish} action has run on all slabs.

We require that dependencies execute slabs sequentially (e.g.~\texttt{TaylorForce[1]} must always execute before \texttt{TaylorForce[2]}), even if preconditions are satisfied out-of-order (this is rare but could happen due to out-of-order I/O completion, for example).  Out-of-order execution is safe but potentially wastes memory since the pipeline effectively becomes wider (more slabs in memory).

\Abacus has several versions of the slab pipeline: notably, on-the-fly group finding and the MPI parallel code introduce a number of new dependencies. These lengthen and widen the pipeline (the pipeline \textit{length} is the number of dependencies a slab must pass through; the pipeline \textit{width} is the number of slabs required to be in memory before the first slab can finish).  Toggling between the multiple versions of the pipeline is as simple as replacing inactive dependencies with dummy preconditions that always pass and ``no-op'' actions that do nothing.  The event loop can remain the same.




\subsection{Insert List and Merge Slabs}
Particles are passed between slabs in the pipeline via the \textit{insert list}.  When a particle drifts from one cell into another, it cannot immediately be inserted into its new location, as slabs are fixed-size arrays, not dynamic lists.  Therefore, when a particle drifts out of a cell, it is removed from the cell and pushed to the insert list.  Then, after slab $i$ and both of its neighbors, $i-1$ and $i+1$, have drifted their particles, the \textit{merge slab} for slab $i$ is constructed, consisting of active particles and any particles from the insert list that belong to slab $i$.  The merge slab is written to state storage (disk or ramdisk), and becomes the input slab for the next time step.

The insert list is a contiguous, fixed-size array, decomposed into \textit{gaps} a few KB in size.  Each thread pushes to its own gap, and when a gap is filled, a new gap is grabbed atomically from the tail of the array.  This allows multi-threaded pushes to the insert list without dynamic memory allocation.

A particle is removed from its old cell by overwriting it with the particle at the tail of that cell.  The ``active'' count of particles in the cell is then decremented, and only active particles are retained during construction of the merge slab.

To ensure that particles will not arrive from 2 or more slabs away, the time step is limited by the maximum velocity in the box, but with Mpc-scale cells, this condition is rarely triggered.  The span of slabs that may drift into a given slab is called the slab width, and is typically 3, as shown in Figure~\ref{fig:pipeline_grid}.

The merge slab is constructed first by condensing any underfilled gaps in the insert list so that the non-empty region is contiguous.  This region is then partitioned such that particles intended for the current merge slab are moved to the end of the array, in cell order.  The construction of the merge slab then consists of ``zippering'' the active particles in the old slab and the new particles in the insert list.  Because both inputs are in cell order, this can be done as an ordered sweep.  An initial counting pass enables multi-threaded copies.

Once the merge slab is constructed, the old slab is no longer needed, and it is released from memory.  To be conservative, we usually allocate a few slabs' worth of memory for the insert list.

\subsection{Memory Footprint}\label{sec:memory_footprint}

\Abacus represents positions and velocities as six 4-byte floats.  The particle ID and flag bits (lightcones, local density estimate, subsample membership) are encoded in an 8 byte integer.  Cell information consists of an 8-byte offset, two 4-byte counts, and velocity statistics for time stepping, taking 32 bytes per cell total.  The far-field data (multipoles and Taylors) are encoded in $(p+1)^2$ floats per cell, or 81 for $p=8$ (in detail, this is $K(K+1)(p+1)^2/2$ complex values per slab).  The \Abacus state therefore consists of 32 bytes per particle and $32 + 4(p+1)^2(K+1)/K$ bytes per cell.  With typical particle-per-cell values of 20 to 120, this is about 35 to 50 bytes per particle.


Within a simulation time step, there are additional allocations. In particular, on-the-fly group finding (Section~\ref{sec:group}) and the parallel code requires some additional memory.  In parallel, every node must allocate space for incoming slabs even though it may not have finished sending its outgoing slabs, leading to temporarily increase memory usage.  In both the single- and multi-node code, the accelerations (3 floats per particle) are an additional large allocation, but are granular at the level of 1 or 2 slabs, meaning the pipeline will only work one or two acceleration slabs ahead until the trailing slab is released.

The peak memory footprint on a node depends on the pipeline width; that is, the number of slabs that must be in memory for the first slab to finish.  In the simplest case of single-node operation, the width is 7 slabs: three slabs must receive forces, which requires 2 position slabs on either side to fill out the near-field radius.  The three central slabs can then kick and drift, allowing the middle slab to receive drifted particles from its neighbors.  This slab can then finish, making the trailing slab eligible for release, and a new slab can begin loading into memory at the leading edge.

In the large parallel simulations of \AbacusSummit, we were able to run group-finding steps of $6912^3$ particles in \ttt{CPD} 1701 on as few as 50 nodes with 512 GiB each (1 GiB---a binary gigabyte---is equal to 1.074 GB).  This is 11.2 TiB of state, or 229 GiB per node.  The state had to obey a strict 256 GiB per-node limit, since it was stored in ramdisk (Section \ref{sec:ramdisk}).  Ephemeral allocations, like the accelerations, group-finding bookkeeping, and manifest slabs, could use the remainder of the 512 GiB, since they were not stored in ramdisk.


\subsection{Parallel Implementation}\label{sec:parallel}
The origin of \Abacus as an ``out-of-core'' code designed to operate on thin slices of a simulation lends itself to parallelization in a 1D toroidal domain decomposition.  Each node hosts a range of slabs, and the slab pipeline proceeds as before.  Eventually, however, the pipeline will stall because the next slab to be loaded is on the next node.  We address this by a rotation of the domain: when the pipeline finishes all processing on its first slab, such that this slab would normally be written out to disk, this marks a division in the dependencies: no slab lower numbered than this finished slab can still affect a slab higher numbered than the finished slab.  Given this cleaving of responsibility, we then send all information on the lower numbered slabs to the next lower node in the torus.  This is by construction exactly the information that that node needs to finish up to the slab adjacent to the one finished.  We note that this structure is not simply a 1D torus, but actually a rotating decomposition: a node ends the step with a different set of slabs than it begins.  

The 1D parallel decomposition is ultimately not as scalable as a 2D or 3D decomposition.  The implementation, however, is much simpler given \Abacus's slab-oriented nature and limits communication overheads.  The main limitation of the 1D decomposition is one of memory: each node must have enough RAM to hold 10--20 \hMpc\ worth of slabs to support group finding.  Platforms like Summit with 512 GB of RAM per node are well-suited to this parallel strategy; we have tested $6912^3$ simulations using 64 nodes successfully, with perfect weak scaling across a factor of 100 in problem size.  Further details of the parallel implementation is given in \citet{Maksimova+2021}. 

\section{Memory, Data, \& Thead Management}\label{sec:memory_and_threads}
\subsection{Overview}
\Abacus manages its memory, data, and thread pools carefully.  We discuss \Abacus's slab allocator and general-purpose allocator, then turn to OpenMP threading and our custom scheduler that increases NUMA locality.  We then discuss state I/O strategies in two contexts: disk-backed storage and ramdisk.  Finally, we discuss management of GPU and I/O thread pools.

\subsection{Memory Allocation}\label{sec:allocation}
Each ``logical slab'' in \Abacus has many different types of associated data, be it particle data like positions and velocities, or cell data like cell offsets and sizes.  Every slab in \Abacus thus has about 20 different associated ``slab types'', such as \texttt{PosSlab}, \texttt{VelSlab}, or \texttt{CellInfoSlab}.  When requesting a slab from the \Abacus ``slab buffer'', one thus specifies both a slab number (an integer) and a slab type (an enum).

The \Abacus slab buffer interface is a high-level wrapper around a low-level ``arena allocator''.  Arenas are thin wrappers around large, contiguous memory allocations.  The arenas manage metadata about whether an allocation is present and how many bytes have been allocated; they also include a few ``guard bytes'' at either end of each slab to help detect out-of-bounds writes.  The details of slab numbers and slab types are abstracted away from the arena allocator, which uses a flattened ``slab ID''.

\Abacus puts a large amount of pressure on the memory allocator (especially with group finding, which allocates per-group temporary workspace).  Memory allocation (\texttt{malloc()} and the like) is well known to be a slow operation on Linux systems.  Indeed, early versions of \Abacus spent 15\% of the total simulation time just calling \texttt{free()} on arenas!  The implementation in the Linux kernel is not particularly optimized for the \Abacus use-case with our mix of small and large allocations and heavy multi-threading.  Furthermore, the kernel memory manager tries to ensure that memory freed in one process is available for allocation in another process; this requires remapping physical pages to new virtual address spaces. This is solving a harder problem than \Abacus needs---we can safely assume that only one (memory-hungry) process is running at a time.

Our approach to reduce kernel memory allocator pressure is twofold: reuse allocations when possible, and use a parallel, user-space \texttt{malloc} implementation.  Implementing arena reuse was fairly straightforward within our allocator: when we discard a slab, we do not always free it but instead mark it as a ``reuse slab''.  The next time the arena allocator receives a allocation request, it first checks if the reuse slab is present and is large enough to satisfy the allocation.  To facilitate a higher hit rate, we over-allocate arenas by a few percent.  To avoid running out of memory, we only retain at most one reuse slab per slab type (in detail, it is implemented as the $(K+1)$-th slab).  In medium-sized simulations (e.g.~the $N=2048^3$, $K=693$ \textit{Euclid} simulation of \citealt{Garrison+2019}), this reduces the number of fresh allocations from $>10^4$ to a few hundred.

We also replaced the built-in GNU allocator with Google's \texttt{tcmalloc} (``thread-cache malloc''), a user-space allocator with high performance under multi-threaded workloads\footnote{\url{https://gperftools.github.io/gperftools/tcmalloc.html}}.  As the name suggests, every thread keeps a cache of recently released memory, such that the allocator can often immediately satisfy small requests out of thread-local cache rather than a central store (thus no synchronization is required).  For large requests, \texttt{tcmalloc} does use a central store, but typically does not release memory back to the kernel.  Thus, all allocations after a short burn-in period can be satisfied in user-space without an expensive kernel call.

\texttt{tcmalloc} has been very successful in handling \Abacus's memory pressure.  When it was first implemented, the 15\% time spent calling \texttt{free()} disappeared, but more surprisingly, it accelerated several independent areas of the code that we did not even realize were affected by background memory management issues.

One risk of a user-space allocator is memory fragmentation.  Memory can fragment in user-space because a new virtual address cannot be assigned to a page until it is released to the kernel.  Therefore, in memory constrained simulations, we release memory to the kernel at a few key moments in the time step, such as after deallocating the sent manifest slabs.  This release is fast, but not totally free, taking about 1 second per time step in the \AbacusSummit simulations.

\subsection{Custom OpenMP Thread Scheduler and NUMA}\label{sec:affinity}
Modern multi-socket platforms often have certain memory banks associated with certain CPU sockets.  All memory remains addressable by all CPUs, but at different rates.  This model is known as ``non-uniform memory access'', or NUMA.

\Abacus is a NUMA-aware code: we try to ensure that CPUs, GPUs, and disks (where applicable) only access memory on their NUMA node.  We accomplish this in two parts: binding threads to cores, and scheduling threads over memory consistently.  The goal of the thread binding is NUMA consistency: that the socket that first touches a region of memory is the only socket that works on that region, as physical memory pages are ordinarily allocated on first touch.  Binding also prevents unnecessary jumping of threads among cores and flushing of caches.  This typically improves \Abacus slab pipeline performance by 20\%.

We schedule threads over $y$-pencils in \Abacus, so by keeping threads within their NUMA nodes, we keep contiguous chunks of slabs in each NUMA node.  For example, with 2 NUMA nodes, each slab is divided in half in physical memory, corresponding to pencils 0 to \ttt{CPD}/2 and \ttt{CPD}/2 to \ttt{CPD}.

The OpenMP thread affinity is controlled via the OpenMP ``places'' mechanism (the \ttt{OMP\_PLACES} and \ttt{OMP\_PROC\_BIND} environment variables).  Each place is a set of one or more cores to which a thread may be bound; the binding parameter controls how to distribute threads to places.  In \Abacus, we typically assign one place per core, so the binding is trivial.  These environment variables are set by the Python wrapper before calling the slab pipeline executable, as OpenMP offers no mechanisms to set these values at runtime.

We implement a custom OpenMP scheduler to schedule threads dynamically within their NUMA node.  OpenMP 4 offers no NUMA-aware scheduler or formal API to implement a custom scheduler, so we build one on top of OpenMP Tasks and atomic addition. The API is a preprocessor macro called \ttt{NUMA\_FOR} that replaces the usual \ttt{for} loop construct.  Upon entering a loop, the loop iterations are divided over the NUMA nodes in proportion the number of threads bound to each NUMA node (computed at program startup).  Each NUMA node tracks a cache-line padded integer containing the next available iteration.  Each OpenMP task atomically captures this value and increments it to determine its next loop iteration.

Because work is distributed from a central ``queue'', scheduling overhead is a concern, notably from the atomic increment and cache-line synchronization.  This approach would not be suitable for ultra-fast loop iterations.  However, at the pencil level, we have not observed any scheduling overhead.  In the \AbacusSummit simulations, this scheduler improved Multipole and Taylor performance by 25\%.

\subsection{I/O Strategies}
When data is not active on the slab pipeline, it must reside in persistent storage, be it disk or ramdisk.  We now discuss strategies for efficient disk I/O, followed by efficient ramdisk usage in Section \ref{sec:ramdisk}.

High-performance I/O on the large, multi-GB state files can be achieved in a number of ways.  One strategy is to maintain two separate filesystems backed by separate media and read from one while writing to the other, and then reverse the roles on the next time step---we refer to this as ``sloshing''.  Sloshing can be beneficial on systems with both high read and write performance on large files.  Another alternative is to assign even-numbered state files to one filesystem and odd-numbered files to another---this is refered to as ``striping''.  Striping can be beneficial when a system prefers a mix of reads and writes, so as to balance the expensive write work across disks, for example.  The optimal choice will depend on the performance and capacity of local disk systems; we sloshed the state and striped the multipoles/Taylors for maximum performance in the \textit{Euclid} simulation of \cite{Garrison+2019}, for example.

Because each \Abacus slab is read and written once per time step, the slab files do not have temporal locality and do not benefit from caching.  To avoid memory pressure from such unnecessary caching, we use Linux Direct I/O, which bypasses the filesystem cache.  However, we have observed that some network filesystems have substantially lower performance with Direct I/O.  The semantics of Direct I/O may be different in these environments; it is possible Direct I/O bypasses some important layers of caching in network filesystems.  Therefore, we turn off Direct I/O except with near-node storage.

\subsection{In-Memory Operation (Ramdisk)}\label{sec:ramdisk}
\Abacus is designed to operate on problem sizes that do not fit into memory by buffering particle data on hard drives and only loading a few slabs into memory at a time.  However, since 2018, GPU hardware (combined with several software engineering improvements in \Abacus) has put the compute rate substantially out of balance with the disk I/O rate.  For our flagship ambitions with \AbacusSummit, it became apparent that even enterprise NVMe drives would struggle to keep up with the \Abacus compute rate.  Rather than alter the foundations of the data model, we instead developed an in-memory version relying on a ramdisk.

In today's Unix systems, by default half of the memory is available as inter-process shared memory, presented as a file system.  We therefore simply write the slabs to this memory instead of to hard disk.  The extra memory required inside the slab pipeline remains in the process memory.  For our parallel application on Summit, this was a reasonable balance.

In principle, this requires very little code change beyond directory path names and scripts to occasionally load or save the shared memory to permanent disk.  However, it would be wasteful to load slabs from the ramdisk to the process memory by duplicating them.  Instead, we use the Linux \code{mmap} function to allow us to access these files directly as slab arenas, thereby saving the memory and making I/O effectively instantaneous.

While this was reasonably quick to implement, we did encounter an unexpected performance impact, finding that mapping and unmapping large portions of memory incurred notable system overheads, of order 10\% of the run time, that we were unable to mitigate.  Nevertheless, this is still far better performance than a hard disk implementation.

Because the ramdisk model exposes slab allocations to a file system path, checkpointing the simulation becomes as simple as executing a file system copy to external storage, such as a network file system.  This checkpoint model is presented in the context of \Abacus on Summit in \cite{2021arXiv210213140G}.

\subsection{GPU \& IO Threads}
Besides OpenMP threads, \Abacus has a GPU thread pool and an I/O thread pool.  The pools are implemented with POSIX \texttt{pthreads} and are bound to cores in a user-customizable manner via the \Abacus parameter file.  We typically assign these threads to the CPU socket to which the corresponding hardware device (GPU or storage drive) is attached.

We set up one GPU work queue per NUMA node; each GPU thread listens on the queue of its NUMA node.  Work units are dispatched to the queue of their central $y$ pencil and will thus remain within their NUMA node, except for potentially at the NUMA boundary.

We typically find it advantageous to give I/O threads their own cores and let each GPU spread its threads over a handful of cores.  This seems to affect I/O and GPU communication at the 20\% level.  The OpenMP threads use the rest of the available cores.  The OpenMP work doesn't scale perfectly with cores in any case, especially in memory-bandwidth-limited operations like the kick and drift, so losing several cores is a worthwhile tradeoff for more GPU performance (especially at late times).  On many-core platforms like Summit (42 cores), this is an even better tradeoff.

\texttt{tcmalloc}, the memory allocator used by \Abacus (Section \ref{sec:allocation}), also has nice NUMA properties.  Since every thread keeps a cache of recently released memory, a thread will likely receive memory that it recently released in response to a new allocation request.  Since \ttt{tcmalloc} usually does not release memory back to the kernel, this means that the memory will still be on the same NUMA node.

\section{On-the-fly Group Finding}\label{sec:group}
\subsection{Overview}
\Abacus is designed for massive simulations where post-processing is expensive---one often does not want to save full particle data from more than a few epochs.  Some data products, such as halo catalogs, we would prefer to have at more epochs than we have full particle outputs, especially for analyses like merger trees.  Thus, on-the-fly analysis is desirable when it will not badly slow down the simulation.  With \Abacus, we have a further requirement: the on-the-fly analysis must be posed in a manner that does not require all particles in memory at once.  In other words, it must be implementable in the slab pipeline.

We have developed an on-the-fly friends-of-friends \citep[FoF,][]{Davis+1985} group finder that is integrated with the \Abacus cell and slab structure.  The FoF group decomposition provides a segmentation of the particles inside of which we do additional local group finding, specifically  the CompaSO halo finder detailed in \cite{Hadzhiyska+2021}.  Here, we will focus on the FoF stage and the slab pipeline aspects.

Aside from generating a useful data product, on-the-fly group finding will have an alternate use in \Abacus: our microstepping (adaptive time stepping) scheme will use these groups to identify regions of small dynamical time that are candidates for sub-cycling.  This will be presented in future work.  Because of this, we seek for group finding to take an amount of time comparable to the rest of the timestep.

\Abacus FoF has the option to limit which particles are eligible to participate in group finding based on a local density estimate (computed on the GPU, see Section \ref{sec:gpu_kernel}).  This allows larger linking lengths to be used without percolating.  The motivation is to set the boundary of the group based on a more precise density estimate, using 10-20 particles rather than the distance to a single nearest neighbor.  Algorithmically, the group finding proceeds as if the ineligible particles were not present.  

\subsection{Algorithm}

Our FoF algorithm consists of three major steps.  First, we perform  open-boundary FoF in every cell, forming \textit{cell groups}, including groups of multiplicity one.  This is trivially parallelizable over cells, and we store the Cartesian bounding box for each cell group.  Second, we search every face, edge, and corner to find pairs of cell groups in neighboring cells that have at least one particle pair below the linking length.  All of these links between cell groups are registered.  Finally, we traverse these graphs of links to find connected sets of cell groups; each such set is a full FoF group, which we call a global group.

When we find cell groups, the cell's particles are reordered so that each group is contiguous.  We then index the cell group by a start index and count.  Global groups then consist of lists of cell groups.  When we want to process further within a global group, we copy these contiguous cell-group particle sets into a buffer so that each global group is contiguous, adjusting the cell-centered positions into group-centered positions.  After any updates, we can scatter the information back into the cell group segments.

Importantly, this algorithm is amenable to a rolling slab pipeline.  Cell group finding can be performed one slab at a time, and the finding of links between cells groups only involves two adjacent slabs at a time.  The step of traversing the links to find the global groups is less local.  We do this by storing a list of links, with each link included twice, once in each orientation.  We find a global group by choosing a cell group, finding all links from it, adding those linked groups to a queue (if they're not already in it), and deleting those links  from the list.  We then repeat that process for every group in the queue.  This produces a list of the cell groups in this global group and ensures that all links involving those cell groups have been deleted from the list of links.  These cell groups are marked as ``closed'', i.e., having been included in a global group.  Finally to complete the concept of how to include this in a rolling pipeline, we perform this activity on a slab basis, starting only from cell groups that have not yet been included in a global group.  For example, if groups can span 10 slabs, then when we search the cell groups in slab 20, we may find examples that involve cell groups in slabs 10 to 30.  Having done this, all of the cell groups in slab 20 will be closed.  When we later search the cell groups in slab 21, all of the cases that link to slab 20 will already have been handled, but we may find new examples that span 21 to 31. 

The group-finding dependencies add significant width to the slab pipeline.  Slabs cannot be released from memory until all groups containing particles in that slab are closed.  FoF, with its extensive percolation, extends the physical pipeline width to 10 \hMpc\ or more, independent of the slab width.  However, the FoF property of local decidability is key for the slab structure of \Abacus, as it allows us to be certain that we have correctly found the groups even when we have access to only a small portion of the simulation volume.  

\subsection{Implementation}

Implementation of this algorithm requires extensive book-keeping and low-level tricks to make it fast.  In particular, we must minimize the $O(N^2)$ work that can happen in FoF, as a singlet particle has to test the separation to every other particle.

In the cell-group-finding step, for cells up to about 70 particles, we perform the simple queue-based $O(N^2)$ FoF algorithm.  One picks a particle, searches all the unaffiliated particles to see if any are close.  If they are, one permutes them to the front of the unaffiliated list.  One then repeats on each next particle in the list, until there are no more affiliated particles that haven't been the search origin.  A trick we use here is to store the particle position and index within the cell as an aligned 4-float set, storing the index as an integer; we then permute these 16-byte objects.  Integers up to about 100M when interpreted bitwise as a float correspond to tiny numbers.  We then compute square distances using AVX instructions in the 4-dimensional space; this is faster than the usual 3-dimensional computation.

\begin{figure}
    \centering
    \includegraphics[width=\columnwidth]{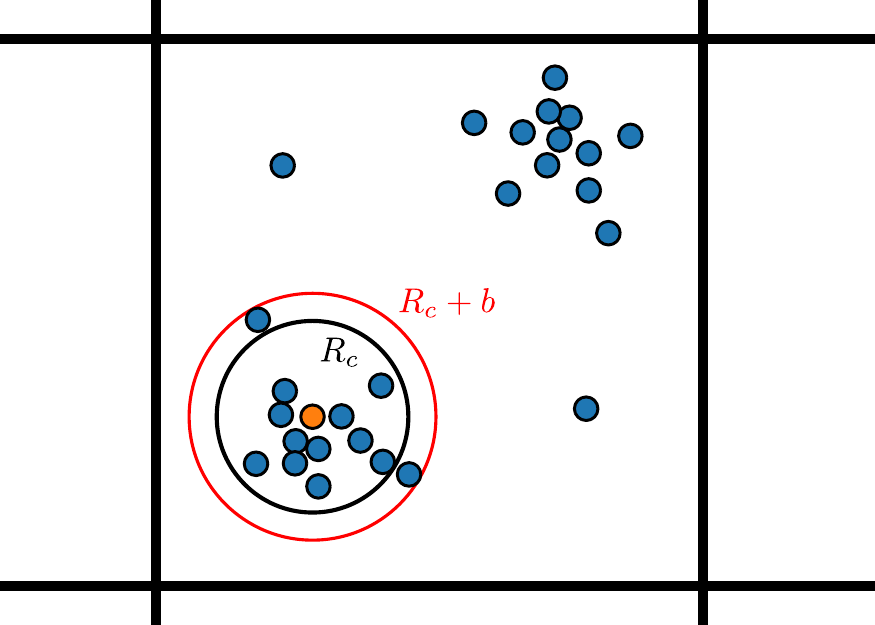}
    \caption{The core-skin partition algorithm to accelerate friends-of-friends (FoF) group finding with linking length $b$.  Particles are partitioned into a core of radius $R_c$, a skin of radius $R_c+b$, and all other particles.  If a FoF link is found from one core particle to another core particle, only distances to core and skin particles must be computed to check for new links. This prunes distant comparisons.  The black lines show the cell grid to illustrate this is applied cell-by-cell.}
    \label{fig:core_skin}
\end{figure}

For cells with more particles, we use similar ideas but we need to reduce the amount of pairwise comparison.  For example, one wants to avoid cases in which two groups in a cell each contain 100 particles, as this would require $100^2$ comparisons.  One could use a tree for this, but we opt for a spherically based decomposition.   For a given first particle, we compute the square distance to all the unaffiliated particles.  We then partition the particles into 3 sets, those within a chosen radius $R_c$, those between $R_c$ and $R_c+b$, and those beyond (Figure \ref{fig:core_skin}).  In this way, we know that when we find a friend particle from the first set, we know that its friends can only be in the first two sets.  As we build up the list of friends, we keep them partitioned into these two sets, so that we search from the friends interior to $R_c$ first.  After we have exhausted these, we have to choose a new center and repeat the distance-based partitioning.  The distance $R_c$ is chosen by experimentation: one wants to include enough particles that one is not partitioning too often, but few enough that the $O(N^2)$ pairwise comparisons are modest.  The algorithm starts with $R_c=2.5b$, but adjusts this value for each repartition.

For the second step of finding links between cell groups, we use the Cartesian bounding boxes to avoid involving any cell groups that are not close enough to the relevant edge, face, or corner.  Further, within a cell group, we isolate those particles that are close to an edge, computing a new bounding box for these.  In this way, we avoid considering most pairs between the two groups.  We also can skip whole pairs of groups if their bounding boxes aren't close enough.  Of course, as soon as a single close pair of particles is found, one can declare this pair of cell groups as a link.

The third step of link traversal is the quickest of the three.  Each cell group is uniquely numbered by the 3-d cell index and then the enumeration of the cell groups.  Each link is two of these cell group numbers; as mentioned above, it is placed in the list twice, in both orders.  We then sort the list on the first number of each link and store the starting indices of each cell.  We mark links for deletion by setting the cell group number to an illegal value; at the end, we sweep through to partition these to the end.

Great attention was paid to the thread parallelization of all of these operations, such that we achieve reasonable load balancing and avoid large numbers of small memory allocations while keeping the bookkeeping and indexing efficient.  Notably, we often decompose by assigning each thread to work on a pencil of constant $x$ and $y$, and use an adaptable storage class to build up lists of unknown length across the cells in a pencil and then between pencils, while providing cell-based indexing into those lists.

\section{Top Level}\label{sec:toplevel}
\subsection{Overview}
We next discuss some of the software ecosystem that surrounds the simulation code proper: the driver code, the inputs (initial conditions) and outputs (particles and halos), analysis tools, and the compilation system.

\subsection{Top-Level Interface}
The top-level \Abacus interface is written in Python.  The Python layer serves several purposes: it provides programmatic and scripting interfaces to set up a new simulation, it parses the parameter file and does preamble work, and it launches the simulation executable.  The main time step loop is actually in the Python layer, since a new process is invoked for every time step.

The executable takes as input the set of slab files on disk, known as the \textit{read state}, and writes out the slab files advanced in time, known as the \textit{write state}.  To advance the time step, the write state directory must be renamed to the read state, and the old read state discarded.  The top-level Python driver takes care of this task, and similar file-related tasks like checking the if the initial conditions are present at the beginning of a simulation, and invoking the IC generator if not.

Checkpoint backups are simple with \Abacus: the state directory \textit{is} the checkpoint.  Thus, making a backup is as simple as copying a directory, usually to some network file system (see \citealt{2021arXiv210213140G} for discussion of \Abacus checkpointing in an HPC context).  The Python layer handles this as well.


The Python layer also sets up some environment variables that the executables will see.  Notably, some OpenMP settings (such as thread affinity, Section \ref{sec:affinity}) have no formal runtime interfaces by which they can be controlled and must be configured with environment variables.



\subsection{Initial Conditions}
The \Abacus initial conditions code, \texttt{zeldovich-PLT}\footnote{\url{https://github.com/abacusorg/zeldovich-PLT}} \citep{Garrison+2016}, is a standard Zel'dovich Approximation \citep{Zeldovich_1970} initial conditions generator with a few modifications.  Just as with \Abacus proper, it is designed to support ``out-of-core'' problems that do not fit in memory.  The large FFT grid is divided into blocks; each block is stored in a separate file on disk.  Planes are loaded into memory by loading the appropriate set of blocks, and the FFT for that plane is performed.  The code uses double-precision internally but can output in either single or double precision.

The code supports optional particle linear theory (PLT) modifications to address the violation of linear theory that occurs on small scales in particle systems \citep{Marcos+2006,Garrison+2016}.  The first modification is to use the growing modes of the particle lattice rather than the continuum system.  This eliminates transients that arise due to the discontinuity in the growth rates between the assumed continuous mass distribution and the actual particle system, and avoids excitation of transverse decaying modes.  The second modification is to rescale the amplitude of the displacement modes in a wave-vector dependent manner to address the fact that the particle system's non-continuum growth rates mean that the power spectrum will not arrive at the value predicted by linear theory at late epochs, at least on small scales.

The code also supports increasing or decreasing the particle count while sampling the same set of modes.  This is achieved via the ``fast-forward'' feature of the Permuted Congruential Random Number Generator \citep[PCG\footnote{\url{https://www.pcg-random.org/}},][]{Oneill_2014}.  This capability is similar to Panphasia \citep{Jenkins_2013}, but \texttt{zeldovich-PLT} does not support changing the box size while holding the modes fixed.


\subsection{Outputs}\label{sec:outputs}
\Abacus has the capability to output a number of data products.  The simplest is a \textit{time slice}, or snapshot, of all particles at a given redshift.  The particles are written in \ttt{CPD} slab files for convenience and as a consequence of the slab pipeline (Section \ref{sec:pipeline})---the \texttt{Output} pipeline dependency is activated when a time slice is requested.

To ensure a synchronous output of positions and velocities, the drift of the step before an output is foreshortened such that the positions land exactly at the target redshift.  The full kick during the output time step ``overshoots'', but the appropriate half-unkick factor is applied during the output.

The time slice outputs are typically written in a bit-packed format called \ttt{packN} ($N$ bytes per particle).  $N$ is either 9 or 14; both versions store the particle kinematic data in 9 bytes, but \ttt{pack14} stores 5 bytes of particle ID information immediately following each nine-tuple of bytes. \ttt{pack9} outputs 8 bytes of particle ID, density, and flag information, but in a separate file.

The positions are stored as 12-bit offsets from cell centers, with 40 bits for particle ID.  Cells usually span a few \hMpc, so a 12 bit offset might pessimistically be 1 \hkpc\ precision.  One may also consider that with a mean PPC of $\mathcal{O}(64)$, cells will have $\mathcal{O}(4)$ particles per dimension initially, and the softening is typically $\mathcal{O}(1/40th)$ of the initial particle spacing, or $\mathcal{O}(1/160th)$ of a cell.  Thus, the quantization is $\mathcal{O}(25)$ times smaller than the softening scale and should not have an impact on any cosmological analysis.

Velocities are also stored in 12 bits, scaled to the maximum box velocity.  This rarely exceeds 6000 km/s, so 12 bits yields 1--2 km/s precision.  These are stored as redshift-space displacements so that the positions and velocities use the same units.

\Abacus particle IDs are assigned from the $(i,j,k)$ index of the initial location in the Lagrangian lattice.  Each component of the index is stored in 16 bits, with one bit spacing between each.  These singleton bits are used for flags.  Interleaving the bits this way reduces the memory footprint when applying compression, as at least half the time they will be next to an identical bit.  The encoding of the Lagrangian index makes it easy to compute the particle displacement from the initial lattice, as well as look up the initial displacement in the initial condition files.


\Abacus can also produce particle light cones.  A light cone is a simulation output in which the box is placed some distance from an imaginary $z=0$ observer and a spherical surface sweeps inwards towards the observer at the speed of light; particles are output when their world lines intersect this surface.  This produces a ``synthetic observation'' that takes into account the finite speed of light.  The light cone outputs are further described in \citet{Maksimova+2021}.

Of course, \Abacus generates many log files.  The primary log file is a verbose record of all slab pipeline operations along with timestamps and copious debugging information.  The other key log file is the timing file: it contains timing breakdowns of each pipeline step and sub-step, along with I/O and GPU performance metrics.  This quickly lets one determine if a simulation is out-of-tune.  Other log files include a log for each I/O thread, and convolution logs.

On-the-fly group finding generates several outputs.  First are the halos: binary records of a few dozen halo properties (mass, velocity dispersion, and so on).  Next are the halo particle subsamples: a user-configurable fraction of the particles in groups, ordered such that group membership can be reconstructed with indexing information from the halo records.  Two such subsamples are allowed (typically one would be small, for HOD work, and one would be larger, for density field work).  The subsamples are selected based on particle ID and are thus consistent across time slice.  These are useful for constructing crude merger trees and as sites of satellite galaxies in the HOD framework.  The particle IDs are output separately from the particle subsample positions and velocities, but in the same order.  Finally, ``tagged'' and ``taggable'' particle are output.  In our hierarchical halo finding scheme, particles are tagged if they are part of the innermost level, called an L2 halo---a halo core.  This allows robust tracking of halos during flybys and mergers.  We describe these further in \citet{Maksimova+2021}.

\subsection{Analysis Tools}\label{analysis}
\Abacus has a few analysis tools for post-processing of massive simulations.  Compared with existing public tools, the primary consideration of our analysis chain is that the particle data may not fit in memory but that we have a guaranteed spatial segmentation of our outputs (slabs).  Therefore, applications that only need a small domain in memory (for example, the two-point correlation function out to a fixed radius) can operate on rolling window with just enough slabs to satisfy the domain requirement.

This analysis pattern is amenable to operation on one node (compared to the MPI design of a package like \textsc{nbodykit}, \citealt{Hand+2018}), given the provenance of \Abacus as a code designed for massive simulations on a single node.  The \Abacus analysis tools are backed by Numba\citep{Lam+2015} and C/C++ via CFFI in performance-intensive regions.  They include a power spectrum code, a correlation function code, and an asynchronous I/O library.



\subsection{Build System}\label{sec:build}
The \Abacus compilation system uses an GNU Autoconf + Make toolchain.  The user runs a \ttt{configure} script which checks for the presence of necessary libraries and sets any compile-time options for the code, such as single or double precision.  Running \ttt{make} will then build the code with those options.

The \ttt{configure} script also outputs a summary of the options the user has selected, an abbreviated version of which is shown in Figure \ref{fig:configure_output}.  This Autoconf-based approach to the build system was inspired by \textsc{Athena} \citep{Stone+2008}.



\begin{figure}
\begin{verbatim}
--------------------------------------------------
Abacus is configured with the following options:
Double precision:                no
Near field max radius:           2
Near field block size:           64
SIMD multipoles:                 avx512
AVX FOF:                         yes
Spherical Overdensity:           yes
GPU directs:                     yes
Compute FOF-scale density:       yes
MPI parallel code:               no
Near-force softening technique:  single_spline
Cache line size:                 64
Page size:                       4096
CXX:                             g++
NUMA-For OpenMP Scheduler:       yes
--------------------------------------------------
\end{verbatim}
\caption{The output of the \Abacus configure script (lightly condensed).}
\label{fig:configure_output}
\end{figure}

\section{\Abacus Hardware}\label{sec:hardware}
\subsection{Overview}
\Abacus was designed to support massive simulations on modest hardware, accessible to a department or lab budget instead of a national supercomputer facility.  As a development environment and proof of concept, we have built a number of such machines in a computer lab at the Center for Astrophysics $|$ Harvard \& Smithsonian.  We will discuss two of the more peculiar requirements of such machines---the disk and the GPUs.  Such purpose-built computers are not necessary---see Section \ref{sec:simulations} for examples of simulations run on standard GPU clusters---but present a compelling opportunity for massive simulations at modest cost.

\subsection{Disk}
For large, single-node simulations, the only ``unusual'' \Abacus hardware requirement is a fast array of disk.  Consider the I/O demands: 32 bytes for particle kinematic data, and about 10 for multipole data (see Section \ref{sec:memory_footprint}).  For a $4096^3$ simulation, using 50 particles per cell and multipole order 8, we thus have 2 TB of particle data and 0.4 TB of multipole data.  To sustain a rate of 20 million particles per second (Mp/s), the total I/O demand (read + write) is thus 1300 MB/s for the particle data.

Our approach is to supply this with hardware RAID (``redundant array of independent disks'') which distributes files over multiple disks to provide some combination of redundancy, performance, and capacity.  We typically use RAID 5 which maximizes performance and capacity while still providing one disk's worth of redundancy (state redundancy is not too important, as it is straightforward to write a simulation checkpoint to another file system).  A single hard drive provides about 200 MB/s under favorable conditions, so with a 10 disk RAID 5 system we could expect 1800 MB/s peak performance (one disk is lost to redundancy).  In practice, we usually achieve 1400 MB/s sustained from 10 disks; at least some of the loss appears to be due to system load (that is, disappears with blocking I/O).  Still, 1400 MB/s is enough to support a compute rate of 20 Mp/s.

Spinning hard drives read and write more slowly towards the center of their platters.  Hard drives consist of several metallic disks (much like small, metal DVDs) with constant areal bit density.  Thus, more bits pass under the read head per rotation on the outer edge than the inner edge.  And since the drives rotate at a fixed rate (typically 7200 RPM), this translates to faster I/O on the outer portion of the platter.

This can be leveraged for better performance.  Hard drives can be ``partitioned'' into logical segments for use by different file systems; this logical partitioning corresponds different physical regions of the hard drive.  By simply creating two partitions per hard drive, one thus segments each drive into a inner, slow partition and an outer, fast partition.  The fast partitions can be linked together in RAID, as can the slow partitions.  This is a convenient split for \Abacus, where we have state files to which we want fast access and output files where performance is not critical.  In practice, the fast partition is consistently 20\% faster than the slow partition, which translates directly to 20\% increase performance in our large, I/O limited sims.  Keeping a ``clean'' partition for the state files also has the benefit of minimizing file fragmentation from small files like logs.

The \Abacus slab I/O pattern of large, bulk reads and writes is quite amenable to RAID with large stripe sizes (the stripe is the atomic unit of RAID operations).  The exception is the convolution: we must hold cross-slab pencils of cells in memory in order to do the $x$-FFT which requires touching all files in small chunks at a time.  Thus, we prefer to use SSDs (solid-state drives) which have nearly no I/O latency and are thus better than hard drives at handling small files (or many files in small chunks).  However, with enough RAM, one can load very large chunks of every multipole file into memory at once, so the cost of using an HDD instead of SSD is not so great.

We note that modern NVMe SSDs (solid-state drives) can provide well over 2 GB/s sustained from a single drive.  However, they are $10\times$ more expensive per GB than HDDs and are only rated for about 1 PB of write---easily achieved in a single large simulation!  The largest drives are still only a few TB, so scaling beyond $4096^3$ in a single node is also not easy.  We have used NVMe drives successfully in more modest, $2048^3$ simulations---those results are presented in \cite{Garrison+2019}.


\subsection{GPUs: Tesla vs GeForce}
In all \Abacus machines that we have built ourselves, we have used consumer-level NVIDIA GeForce cards instead of the HPC-marketed NVIDIA Tesla cards.  The Tesla cards typically have larger memory and vastly superior double-precision performance, but our use of cell-centered particle coordinates ensures that we do not need double precision.  The pencil-on-pencil data model (Section \ref{sec:gpu_data_model}) also does not benefit directly from increased memory capacity.  The price difference is considerable: a Tesla A40 card costs thousands of dollars, while the GeForce RTX 3080 (based on a similar, but not identical, chip) retails at \$700.  The single-precision performance of an A40 is about 50\% better, but the price-to-performance ratio is far worse.

\section{Notable Simulations}\label{sec:simulations}
\Abacus has been used to run a number of large simulations on computer hardware ranging from purpose-built, single node machines (Section \ref{sec:hardware}), to university GPU clusters, to national supercomputer facilities.  We highlight some lessons from each.

In 2016, we used a single, purpose-built node to run a simulation of $5120^3$ particles (130 billion) in a $250h^{-1}\ \mathrm{Mpc}$ box, for a particle mass of $1\times10^7h^{-1} \Msun$.  The simulation, presented in \cite{Zhang+2019}, was designed to explore the detectability of the clustering of the first galaxies with a JWST 13 arcmin deep-field survey.  The simulation was evolved from $z=200$ to $8$ in eight weeks on the \ttt{franklin} hardware (dual Xeon Sandy Bridge, dual NVIDIA GeForce 980 Ti).  In our analysis, we found that the extreme bias factors (5--30) of massive halos at this epoch lend themselves to detection of clustering with only 500--1000 objects, assuming that the detected galaxies occupy the most massive halos.

Our next large simulation on \ttt{franklin} followed shortly thereafter: a more traditional, BAO-oriented simulation dubbed ``FigBox'' of $4096^3$ particles in a $3.2\hGpc$ box, for a particle mass of $4\times10^{10}\hMsun$.  After the JWST simulation, it was apparent that the GPU pencil construction model was a weakness, as it involved two copies: packing the particles from slabs into pencils, and from pencils into pinned memory for staging to the GPU.  Large copies are expensive operations, particularly on Intel platforms where one core can only pull 10 GB/s of bandwidth from main memory.  Thus, we introduced the ``deferred copy'' GPU pencil model described in Section \ref{sec:gpu_data_model}.  The salient part is that a \ttt{PencilPlan} is constructed for each source and sink pencil that contains indexing information but waits to copy any particles until the GPU work unit comes up for execution.  At that point, the particles are packed directly from the slabs into the pinned memory.

This model was very successful, with the overall CPU work running about 30\% faster.  However, the disks could only supply 22 Mp/s, so the wall-clock time to completion was still about 8 weeks.

FigBox was an important testing ground for group finding (Section \ref{sec:group}).  With its large volume, it finds rare peaks and filaments in the cosmic density field that might be missed in a smaller box, and thus helps us understand the percolation properties of various algorithms.  This is particularly important for on-the-fly group finding, where the largest filament sets the number of slabs we must hold in memory, and thus the requisite amount of RAM per node.  A FoF linking length of 0.2, for example, finds a $20\hMpc$ group that is actually a string of 9 or 10 visually obvious halos embedded in a filament.  Our density eligibility criterion now mitigates such percolation.

The first cluster port of \Abacus was to the University of Arizona's El Gato.  In 2016, El Gato was used to run the Abacus Cosmos simulations of \cite{Garrison+2018}, with 150 simulations of 3 billion particles each.  Each simulation was run on one node, enabled by the large RAM per node.  We implemented many improvements for operation in cluster environments, such as build system enhancements and job scheduler interaction.  The slowness of the \ttt{memcpy} incurred by reading files from ramdisk, as opposed to mapping them, became abundantly clear as well (Section \ref{sec:ramdisk}).  Abacus Cosmos has been used in many analyses, such as
\cite{2018MNRAS.478.1866H,
2018MNRAS.478.2019Y,
2019MNRAS.484..989W,
2019MNRAS.485.2407G,
2019MNRAS.486..708Y,
2019MNRAS.490.2606W,
2019MNRAS.490.2718D,
2020MNRAS.491.3061S,
2020ApJ...889..151N,
2020MNRAS.492.2872W,
2020MNRAS.493.5506H,
2020MNRAS.493.5551Y,
2020PhRvD.101l3520P,
2021MNRAS.502.3582Y}.

In 2018, \Abacus was used to produce a high-fidelity realization of the \textit{Euclid} code comparison simulation \citep{Garrison+2019}.  We demonstrated \Abacus's performance on commodity hardware, running the $2048^3$ simulation with $1.2\times 10^{10} \hMsun$ particle mass in 107 hours on a single node (dubbed \texttt{hal}), using two Intel Xeon CPUs, two NVIDIA 1080 Ti GPUs, and fast disk.  We demonstrated \Abacus's force accuracy---many of these tests are repeated in this work (Section~\ref{sec:force_accuracy})---and relative insensitivity to time step, except for at the smallest scales, in halo cores.  \Abacus was shown to reproduce the linear-theory solution to better than 0.01\%.

The most ambitious \Abacus simulations to date are the \AbacusSummit simulations\footnote{\url{https://abacussummit.readthedocs.io/}}, run on the Summit supercomputer (recently \#1 on the Top500 list\footnote{\url{https://top500.org/lists/top500/2019/11/}}).  Consisting of 60 trillion particles spanning 97 cosmologies, the planning, execution, and data management were all substantial challenges, which are detailed in \cite{Maksimova+2021}.  The Summit node architecture, with its high memory bandwidth, 44 CPU cores, and 6 NVIDIA V100 GPUs was well-suited to \Abacus, yielding 70 million particle updates per second in unclustered states, and 45 million at the terminal redshift of $z=0.1$.

As part of that time allocation, \Abacus was used to run several $6144^3$ particle high-redshift simulations in small boxes (20, 80 and 300 \hMpc) for reionization studies.  Several scale-free simulations of varying spectral index ($n_s=-1.5$, $-2$, $-2.25$, $-2.5$) in $4096^3$ and $6144^3$ were run as well, adopting the same normalization and output conventions as \cite{Joyce+2020}.  Analysis of these simulations will be presented in future work.

\section{Summary}\label{sec:summary}
The \Abacus $N$-body code achieves simultaneous high accuracy and performance through a new mathematical method for computing the long-range gravitational force---a convolution over a static multipole mesh, imparting a disjoint near-field and far-field.  Coupled with GPUs for massive acceleration of the near-field force, this method is very efficient at solving the $N$-body simulations required by modern large-scale structure surveys---large, moderately-clustered volumes.

The \Abacus far-field method is fast even at high multipole order---most \Abacus simulations to date, including the \AbacusSummit suite, have used order 8.  With low-level CPU optimization, we achieve a seven-fold speedup over the naive implementation of the multipoles, processing more than 50 million particles per second per core while yielding typical force errors of $10^{-5}$ to $10^{-6}$.  In the near-lattice configuration of typical cosmological initial conditions, the power in the error is at least 6 orders of magnitude below the power in the displacements.

With the \AbacusSummit suite \citep{Maksimova+2021}, \Abacus has been deployed to generate cosmological sims at a scale never achieved before---60 trillion particles spanning 97 cosmologies, featuring 137 simulations with a base mass resolution of $2\times10^9\hMsun$ in a $2\hGpc$ box, each costing 1800 Summit node-hours.   These simulations were designed to meet and exceed the cosmological simulation requirements of DESI. 

As computer hardware evolves in fulfillment of the needs of major applications, such as machine learning, graphics rendering, and data analytics, scientific computing must evolve to exploit these new opportunities.  \Abacus does so at multiple levels, coupling high-level code design, mathematical methods, and low-level optimization to find order-of-magnitude opportunities in the $N$-body problem.  The domain decomposition into slabs of cells allows the computation to be organized as an event-driven pipeline of actions and dependencies, allowing the maximum amount of computation to proceed while reducing the in-memory footprint to a narrow slice of the simulation volume.  This ``rolling window'' model couples with the mathematical property of a disjoint near-field to allow the exact near-field component to be computed from a strictly bounded domain.  The computation is then accelerated by careful packaging of these cells of particles into pencils that can be efficiently processed by the GPU.  The CPU, meanwhile, handles the cell-centered multipoles with SIMD-vectorized kernels.  \Abacus is highly optimized to work on today's hardware while retaining general design principles, such as massive parallelism in the near-field computation, that will scale to many future generations of computer hardware.


\acknowledgments
The authors extend their gratitude to Marc Metchnik, whose Ph.D.~thesis initiated the \Abacus project.  We would additionally like to thank Salman Habib, David Spergel, and David Weinberg for helpful conversations, the referee and Volker Springel for constructive comments, and Lisa Catella for many years of administrative support.

This work has been supported by NSF AST-1313285 and DOE-SC0013718, as well as by Harvard University startup funds.
DJE is supported in part as a Simons Foundation investigator. 
NAM was supported in part as a NSF Graduate Research Fellow.  
LHG is supported by the Center for Computational Astrophysics at the Flatiron Institute, which is supported by the Simons Foundation.  
PAP was supported by NSF AST-1312699.
The \textsc{AbacusCosmos} simulations were run on the El Gato supercomputer at the University of Arizona, supported by grant 1228509 from the NSF.

The \AbacusSummit simulations used resources of the Oak Ridge Leadership Computing Facility, which is a DOE Office of Science User Facility supported under Contract DE-AC05-00OR22725.   The \AbacusSummit simulations have been supported by OLCF projects AST135 and AST145, the latter through the Department of Energy ALCC program.

%


\software{Astropy \citep{Astropy_2018},
            NumPy \citep{van_der_Walt+2011},
            SciPy \citep{Virtanen+2020},
            Numba \citep{Lam+2015},
            CUDA \citep{Nickolls+2008},
            Intel TBB \citep{Reinders_2007},
            matplotlib \citep{Hunter_2007},
            FFTW3 \citep{FFTW05},
            Corrfunc \citep{Sinha_Garrison_2019,Sinha_Garrison_2020},
}
          
\section*{Data availability}
Clustering measurements and other summary statistics used in this work are available upon request.  The underlying simulation data is much larger, but may also be made available upon reasonable request.  Data related to the \AbacusSummit suite is available at \url{https://abacussummit.readthedocs.io/en/latest/data-access.html}.






\bibliography{biblio}{}
\bibliographystyle{aasjournal}



\end{document}